\documentclass[lettersize,journal]{IEEEtran}
\usepackage{amsmath,amsfonts,amsthm,amssymb}
\usepackage{mathrsfs}
\usepackage[linesnumbered,ruled]{algorithm2e}
\usepackage{array}
\usepackage{textcomp}
\usepackage{stfloats}
\usepackage{url}
\usepackage{verbatim}
\usepackage{graphicx}
\usepackage{cite}
\usepackage{nicematrix}
\usepackage{lipsum}
\usepackage{tikz}
\usepackage{pdfpages}
\usepackage{bm}
\usepackage{booktabs}
\usepackage{multirow}
\usepackage{enumitem}
\usepackage{subcaption}
\usepackage{pifont}
\usepackage{xcolor}
\usepackage{colortbl}

\usepackage{multirow}
\usepackage{makecell}
\usepackage{float}
\usepackage{authblk}
\usepackage{url}

\usepackage{colortbl}  
\usepackage{wasysym}

\definecolor{gray0}{gray}{0.9}
\newtheorem{definition}{Definition}[section]

\def \toolname{WGLE\xspace}

\begin{document}

\title{\toolname: Backdoor-free and Multi-bit Black-box Watermarking for Graph Neural Networks}

\author{Tingzhi Li, Xuefeng Liu, Jing Lei, Xingang Zhang
\thanks{Tingzhi Li is with Nanyang Normal University, Nanyang 473000, China. (e-mail: tingzhi.li@nynu.edu.cn)}
\thanks{Xuefeng Liu and Jing Lei are with Xidian University, Xi'an 710000, China. (e-mail: liuxf@mail.xidian.edu.cn, leijing@xidian.edu.cn)}
\thanks{Xingang Zhang is corresponding author and with Nanyang Normal University, Nanyang 473000, China. (e-mail: zxg@nynu.edu.cn)}
}

\markboth{Journal of \LaTeX\ Class Files,~Vol.~14, No.~8, August~2021}%
{Shell \MakeLowercase{\textit{et al.}}: A Sample Article Using IEEEtran.cls for IEEE Journals}

\IEEEpubid{0000--0000/00\$00.00~\copyright~2021 IEEE}

\maketitle

\begin{abstract}


Graph Neural Networks (GNNs) are increasingly deployed in real-world applications, making ownership verification critical to protect their intellectual property against model theft. Fingerprinting and black-box watermarking are two main methods. However, the former relies on determining model similarity, which is computationally expensive and prone to ownership collisions after model post-processing. The latter embeds backdoors, exposing watermarked models to the risk of backdoor attacks. Moreover, both previous methods enable ownership verification but do not convey additional information about the copy model. If the owner has multiple models, each model requires a distinct trigger graph.

To address these challenges, this paper proposes \toolname, a novel black-box watermarking paradigm for GNNs that enables embedding the multi-bit string in GNN models without using backdoors. \toolname builds on a key insight we term Layer-wise Distance Difference on an Edge (LDDE), which quantifies the difference between the feature distance and the prediction distance of two connected nodes in a graph.  
By assigning unique LDDE values to the edges and employing the LDDE sequence as the watermark, \toolname supports multi-bit capacity without relying on backdoor mechanisms. We evaluate \toolname on six public datasets across six mainstream GNN architectures, and compare \toolname with state-of-the-art GNN watermarking and fingerprinting methods.
\toolname achieves 100\% ownership verification accuracy, with an average fidelity degradation of only 1.41\%. Additionally, \toolname exhibits robust resilience against potential attacks. The code is available in the repository.\footnote{\url{https://github.com/tingzhi-li/WGLE}}

\end{abstract}

\IEEEpeerreviewmaketitle
\section{Introduction}
\label{Introduction}
\IEEEPARstart{D}{espite} the growing deployment of Graph Neural Networks (GNNs) in graph-related applications~\cite{harl2020explainable,tan2019deep,wang2023doitrust}, the problem of intellectual property (IP) protection for GNNs remains fundamentally under-explored. Existing approaches rely on ownership verification and fall into two dominant paradigms: backdoor-based black-box watermarking and model fingerprinting. 

However, these two types of methods have inherent limitations when applied in real-world scenarios. In particular, 
backdoor-based black-box watermarking methods~\cite{zhao2021watermarking,xu2023watermarking} embed trigger-induced misclassification behaviors into GNNs, enabling ownership verification through abnormal predictions. While effective, such approaches inherently introduce exploitable backdoors, posing severe security risks once exposed. 
In contrast, model fingerprinting methods~\cite{waheed2024grove,you2024gnnfingers,zhou2024revisiting} avoid explicit backdoors by relying on model similarity comparison, but they suffer from high computational overhead, vulnerability to post-processing (e.g., fine-tuning and pruning), and non-negligible false positive rates~\cite{pan2022metav, liu2024false}. More critically, almost all existing GNN ownership verification methods are limited to zero-bit capacity, requiring a unique trigger graph for each protected model. This limitation leads to significant verification cost, poor scalability, and susceptibility to ownership ambiguity attacks.

\IEEEpubidadjcol

\noindent \textbf{Problems.} These shortcomings reveal a fundamental gap:
\textit{Can GNN ownership be verified in a black-box setting without embedding backdoors, while supporting scalable multi-bit identification?}
Our work aims to fill this gap by systematically investigating the following problems:

\begin{enumerate}[label=\textbf{P\arabic*.}, ref=\textbf{P\arabic*}, leftmargin=*]
    \item What intrinsic properties of GNNs, if any, can be leveraged as stable and controllable watermark carriers instead of backdoor under black-box access?
    \item Is it possible to achieve black-box ownership verification for GNNs without embedding any backdoor behaviors that could be exploited by malicious users?
    \item Can a single trigger graph support multi-bit watermarking, i.e., enable the owner to extract distinct watermarks from multiple watermarked GNNs using the same graph?
\end{enumerate}

\noindent Addressing these problems requires rethinking how information is leaked through GNN predictions beyond individual sample predictions.

\noindent \textbf{Challenges.} Designing a backdoor-free, multi-bit black-box watermarking scheme for GNNs is challenging due to several fundamental obstacles shown as follows:
\begin{itemize}[leftmargin=*]
    \item \textit{Severe scarcity of controllable carriers in black-box settings.} In black-box access, only model predictions are observable, resulting in a severe scarcity of usable watermark carriers.
    \item \textit{Coupling between node predictions and graph topology.} In GNNs, predictions are not independent across samples. Message passing entangles node features, labels, and topology. While the prior method for traditional models (i.e., models for text, images) leverage specific entropy values as carriers instead of backdoor~\cite{li2022untargeted}, GNN predictions are highly sensitive to minor local perturbations, making it difficult to embed stable watermarks by this strategy~\cite{dai2024comprehensive}.
    \item \textit{Multi-bit capacity under a single trigger graph.} One approach to achieve multi-bit capacity in black-box settings for traditional models is using multiple independent triggers~\cite{chen2019blackmarks}. However, this strategy still relies on backdoors and yields only limited watermark capacity, making it unsuitable for many watermarked models.
\end{itemize}
These challenges collectively explain why existing GNN watermarking methods predominantly rely on backdoor-based approaches, and why directly applying watermarking methods in traditional models to GNNs remains non-trivial.

\noindent \textbf{Motivation and Key Insight.} Our key motivation stems from a fundamental observation:
\textit{A black-box GNN reveals not only individual node predictions, but also implicit relational information between nodes.}

\begin{figure}[!t]
    \centering
    \includegraphics[width=1.0\linewidth]{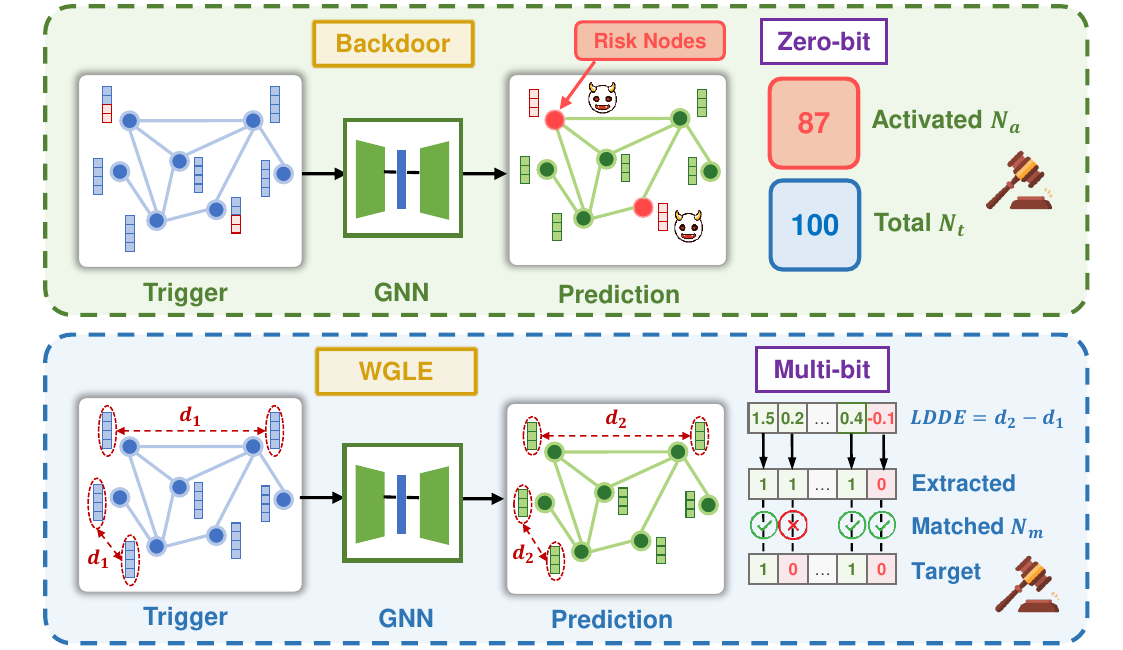}
    \caption{The differences between \toolname and the previous GNN black-box watermarking. Previous methods use backdoors, with ownership verification relying on misclassification of trigger samples. \toolname uses LDDE as watermarks, enabling verification without potential security risks.}
    \label{fig:intro}
    \vspace{-3mm}
\end{figure}

As presented in Figure~\ref{fig:intro}, prior black-box watermarking methods for GNNs focus exclusively on individual sample predictions.
GNNs inherently encode relational dependencies through message passing. These dependencies are reflected in the relative distances between connected nodes in both feature space and prediction space.
Based on this observation, we identify a relational leakage channel in black-box access for GNNs, which we formalize as \textit{Layer-wise Distance Difference on an Edge (LDDE)}-the difference between prediction distance and feature distance for connected node pairs.
Crucially, LDDE exhibits three desirable properties:
\begin{enumerate}[leftmargin=*]
    \item It exhibits broad compatibility with diverse GNN architectures and can be calculated under black-box access.
    \item LDDE signs for certain edges can be precisely controlled to turn desired signs via model fine-tuning.
    \item Such modifications have minimal impact on the model's primary task performance.
\end{enumerate}
These properties make LDDE an ideal watermark carrier (\textbf{P1}), enabling ownership verification without inducing backdoor behaviors (details in Section~\ref{Insight}).

As illustrated in Figure~\ref{fig:intro}, backdoor-based watermarking methods rely on the backdoor sample activation rate, which is a single-bit indicator determined solely by whether this rate exceeds a predefined threshold. Some samples are misclassified into the specific class due to backdoor-induced behavior.
In contrast, there exists no pattern in features that deterministically leads to the specified classification in \toolname. Each node retains its original features, while the edges between these nodes carry specific LDDE values that serve as the watermark (\textbf{P2}).
Each edge can be assigned a distinct LDDE value, and the signs of these LDDE values form a sequence that is subsequently mapped to a multi-bit binary string (\textbf{P3}).
During verification, the verifier extracts a multi-bit watermark string from the predictions of the suspect model on the trigger graph. Each watermarked model has a unique watermark string, yet all models are queried using the same trigger graph.

Considering practical deployment requirements, we propose two watermark embedding settings: Setting that the Training graph is Available (\textbf{S.T.A.}), where the owner utilizes the original training graph directly as the trigger graph for watermark embedding and verification. Setting that the Training graph is Missing (\textbf{S.T.M.}), where the owner generates a graph as the trigger graph and employs data-free adversarial knowledge distillation to embed the watermark without access to the training graph. This setting is particularly suitable for ownership transfer scenarios where the new owner possesses only the GNN model but lacks the training data.

\noindent \textbf{Evaluations.} 
We evaluate \toolname and state-of-the-art GNN black-box watermarking (WGB~\cite{xu2023watermarking}) and fingerprinting methods (RBOVG~\cite{zhou2024revisiting}) across six mainstream GNN architectures and six real-world datasets. 
The experimental results show that \toolname achieves 100\% ownership verification accuracy while being backdoor-free and supporting multi-bit capacity.
Regarding fidelity, \toolname incurs only a 2.37\% drop in mean test accuracy under \textit{S.T.A.} and a 0.45\% drop under \textit{S.T.M.}, compared to 1.21\% for WGB and 3.13\% for RBOVG. 
\toolname maintains a zero false positive rate across all experiments, whereas RBOVG exhibits non-zero false positive rates after pruning or fine-tuning. \toolname in \textit{S.T.A.} is robust against model extraction attacks, whereas WGB struggles with this attack. 

\noindent \textbf{Contributions.}
This paper proposes \toolname, a novel GNN black-box watermarking paradigm for GNN ownership verification. \toolname inherits the advantages of low overhead and strong robustness of GNN black-box watermarking methods, and eliminates the risks of using backdoor and supporting multi-bit capacity.
The contributions of our paper are summarized as follows:
\begin{itemize}[leftmargin=*]
    \item \textit{A new security perspective on GNN watermarking.} We identify and exploit a previously overlooked relational leakage channel (LDDE) in black-box access for GNNs, and demonstrate it suitable as black-box watermarking carrier.     
    \item \textit{A backdoor-free, multi-bit black-box watermarking framework for GNNs.} We propose \toolname, a novel GNN black-box watermarking paradigm that does not use a backdoor and enables the verifier to extract distinct multi-bit watermark strings from suspect models via a single trigger graph.
    \item \textit{Comprehensive embedding settings for practical deployment.} We design two complementary watermark embedding modes: \textit{S.T.A.} for scenarios where the training graph is available, and \textit{S.T.M.} for ownership transfer scenarios where the training graph is missing.
    \item \textit{Comprehensive security evaluation.} Experiments across six datasets and six GNN architectures present that \toolname achieves perfect ownership verification accuracy, zero false positives, high fidelity, and strong robustness against common post-processing attacks, including pruning, fine-tuning, watermark overwriting, and model extraction attacks.
\end{itemize}

\section{Background}
\label{Background}
\subsection{Graph Neural Networks}
Graph neural networks (GNNs) are widely adopted for various graph data processing tasks~\cite{harl2020explainable,lv2019auto,li2021braingnn,tan2019deep,fan2019graph}. GNNs create vector representations (embeddings) for nodes by iteratively gathering information from their local neighborhoods. At the $l$-th layer, the embedding $h_v^l$ of node $v$ is computed by aggregating the embeddings $h_u^{l-1}$ of its neighbors $u \in \mathcal{N}(v)$:
\[
h_v^l = \text{AGG}\left(h_v^{l-1}, \left\{ \text{MSG}\left(h_v^{l-1}, h_u^{l-1}\right) \mid u \in \mathcal{N}(v) \right\} \right),
\]
\noindent where $h_v^l$ denotes the embedding of node $v$ at layer $l$, with $h_v^0$ initialized from the node feature $\mathbf{x}_v$. The message function $\text{MSG}$ computes messages exchanged between node $v$ and its neighbors $u$ based on their embeddings at layer $l-1$. The aggregation function $\text{AGG}$ combines information from the last layer’s embedding $h_v^{l-1}$ and the message generated by $MSG$.

Due to the prevalence and broader applicability of node-level tasks, this paper focuses on node-level GNNs, consistent with previous work~\cite{defazio2019adversarial,wu2022model,shen2022model}. A node-level GNN takes a graph as input and outputs predictions for each node.
GNNs can follow two training paradigms: (1) \textit{Transductive}, where the training and test graphs share the same structure but focus on different nodes; and (2) \textit{Inductive}, where the model is trained on one graph and evaluated on an unseen graph.

\subsection{Model Ownership Verification}
Model ownership verification determines whether the suspect model $\mathcal{M}_s$ is a copy of the protected model $\mathcal{M}_p$, formalized as a binary decision problem defined as
\[
\mathcal{V}(\mathbf{b}, \mathcal{M}_s, \tau) = \mathbb{I}\{\text{SIM}(\mathbf{b}, \mathcal{F}_{\text{ext}}(\mathcal{M}_s)) > \tau\}
\]
\noindent where $\mathbf{b}$ is the IP identifier (e.g., watermark or fingerprint). $\mathcal{F}_{\text{ext}}(\cdot)$ is the extraction function that extracts the IP identifier from the suspect model $\mathcal{M}_s$. 
$\text{SIM}(\cdot,\cdot)$ is a similarity metric. $\tau$ is the verification threshold. If the similarity exceeds $\tau$, the suspect model is deemed to contain the IP identifier, and thus an copy of the protected model.
Existing IP protection methods can be categorized into three types: (1) white-box watermarking, (2) black-box watermarking, and (3) fingerprinting. 

White-box watermarking methods embed a message in the internal parameters of the watermarked model~\cite{Uchida2017}. 
It requires full access to the internal parameters of the suspect model during ownership verification, which is often impractical in practice. 
Black-box watermarking methods embed a backdoor in the watermarked model~\cite{adi2018turning}. 
The suspect model is verified as a copy if it can be triggered by backdoor samples.
Black-box watermarking methods enable ownership verification in settings where the verifier accesses only the suspect model's input-output behavior via its API, without knowledge of internal parameters. However, these methods pose security risks. Malicious users can exploit the exposed backdoor to focus the watermarked model to produce desired outputs~\cite{zhao2021watermarking,xu2023watermarking}.
Fingerprinting offers an alternative to watermarking for model ownership verification by comparing model similarity on intrinsic properties (i.e. fingerprints)~\cite{waheed2024grove,you2024gnnfingers}. Independently trained models exhibit distinct fingerprints, whereas a copy model shares a highly similar fingerprint with the protected model. 
Fingerprinting methods aim to train a classifier to distinguish independently trained models and copies of the protected model. 


\section{Problem Statement}
\label{ProblemStatement}
\subsection{System and Threat Model}

\begin{figure}
    \centering
    \includegraphics[width=1.0\linewidth]{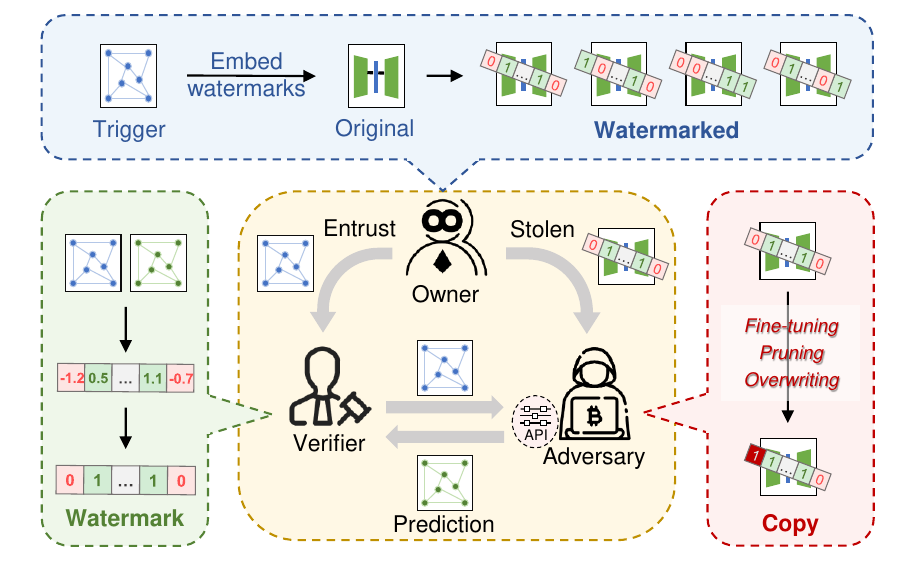}
    \caption{The system model of \toolname. The owner uses the trigger graph to embed watermarks in the original model, producing watermarked models. The adversary steals one of the watermarked models and may post-process the copy model. The verifier uses the trigger graph to query the suspect model's API and extracts the watermark from the predictions.} 
    \label{fig:system}
    \vspace{-3mm}
\end{figure}

Figure\ref{fig:system} describes three entities in \toolname: the original model owner, the adversary, and the verifier. The verifier can be the model owner or a trusted third-party with the trigger graph and watermarks provided by the owner. 

\noindent \textbf{Original Model Owner.}
The owner has the original GNN models, with or without the original training graph (\textit{S.T.A./S.T.M.}). The owner's objective is to embed a watermark into the owned GNNs with low overhead. The owner can directly take the training graph as the trigger graph in \textit{S.T.A.} or generate a graph as the trigger graph in \textit{S.T.M.}. When a client requests a model, the owner generates a unique watermark and embeds it into the original model. It is noticed that all watermarked models share the same trigger graph.

\noindent \textbf{Adversary.}
The adversary is an IP infringer who has an unauthorized copy of the watermarked model. For example, the adversary can steal the entire model parameters through insider leaks or server breaches, or train a surrogate model via model extraction attacks. The adversary attempts to deploy the copy model and provide API access services while evading potential ownership verification. For this purpose, the adversary can reprocess the watermarked model using techniques such as fine-tuning~\cite{church2021emerging}, pruning~\cite{han2015learning}, or watermark overwriting~\cite{lao2022identification}.

\noindent \textbf{Verifier.}
Holding the trigger graph and the watermarks, the verifier aims to verify the ownership of the suspect model. In practice, ownership verification is typically performed in a black-box setting, where the verifier can only interact with the suspect model via its API~\cite{adi2018turning, sun2023deep}. 
The verifier uses the trigger graph to query the suspect model and analyzes the predictions to determine whether the suspect model is an unauthorized copy. 
As mentioned above, the same trigger graph is used for all watermarked models, but can be extracted for respective watermarks from different model outputs. Without the necessary multiple distinct trigger graphs, ownership verification can be completed with just one query to the suspect model. Compared to zero-bit watermarks, multi-bit watermarks are more difficult to forge.

\subsection{Requirements for Ownership Verification.}
\label{requirements}
\toolname is a novel GNN black-box watermarking paradigm that satisfies all the following requirements:
\begin{enumerate}[label=\textbf{R\arabic*.}, ref=\textbf{R\arabic*}, start=0, leftmargin=*]
    \item \textit{Effectiveness and Fidelity:} Effectiveness requires that the scheme reliably distinguishes unauthorized copies from independently trained models. Fidelity requires that the watermarked model preserve a performance comparable to the original model on the primary tasks.
    \item \textit{Backdoor-free:} The watermarked model is backdoor-free, eliminating the potential threat of backdoor attacks.
    \item \textit{Multi-bit:} During ownership verification, the verifier extracts a multi-bit string (i.e., the watermark) from the suspect model. Multi-bit capacity permits the extraction of unique watermarks from various watermarked models by the same trigger graph.
    \item \textit{Robustness:} Ownership verification should remain effective even if the watermarked model is post-processed, such as pruning or fine-tuning.
    \item \textit{Low Overhead:} The watermark embedding process incurs a low computational overhead.
\end{enumerate}

\noindent In these requirements, \textbf{R0} is a basic requirement and must first be satisfied by all ownership verification methods. 

\begin{table}[!t]
\centering
\caption{Comparison of \toolname with previous methods. ``\textbf{R1.}'', ``\textbf{R2.}'', ``\textbf{R3.}'', and ``\textbf{R4.}'' respectively denote backdoor-free, multi-bit, strong robustness, and low overhead. ``\ding{108}'' ``\Circle'' ``\LEFTcircle'' denote full compliance, non-compliance, and  partial fulfillment with room for improvement, respectively. }
\label{tab:compare}
\setlength{\tabcolsep}{8pt}{
\resizebox{\columnwidth}{!}{
\begin{NiceTabular}{c|l|cccc}
\toprule
Category                                & Method                              & \textbf{R1.} & \textbf{R2.} & \textbf{R3.} & \textbf{R4.} \\
\midrule
\multirow{3}{*}{Fingerprinting}         & GrOVe~\cite{waheed2024grove}      & \ding{108}           & \Circle        & \Circle         & \Circle           \\
                                        & GNNFingers~\cite{you2024gnnfingers} & \ding{108}           & \Circle        & \Circle         & \Circle           \\
                                        & RBOVG~\cite{zhou2024revisiting}    & \ding{108}           & \Circle        & \LEFTcircle       & \Circle           \\
\midrule
\multirow{3}{*}{\begin{tabular}[c]{@{}c@{}}Black-box \\ watermarking\end{tabular}} & WGRG~\cite{zhao2021watermarking}   & \Circle            & \Circle        & \ding{108}        & \ding{108}          \\
                                        & WGB~\cite{xu2023watermarking}       & \Circle            & \Circle        & \ding{108}        & \ding{108}          \\
\cmidrule{2-6}
                                        & \toolname                           & \ding{108}           & \ding{108}       & \ding{108}        & \ding{108}
                                            \\         
\bottomrule
\end{NiceTabular}%
}}
\end{table}
\vspace{1mm}

\subsection{Limitations of Prior Work}
\label{sec3.4}
Table~\ref{tab:compare} compares \toolname with existing methods. All works satisfy \textbf{R0}, but previous works lack complete consideration of \textbf{R1}, \textbf{R2}, \textbf{R3}, and \textbf{R4} (introduced in Section~\ref{requirements}). 

\noindent \textbf{GNN Black-box Watermarking.} 
WGRG~\cite{zhao2021watermarking} is the first GNN black-box watermarking method. WGB~\cite{xu2023watermarking} further extends this line of work. They construct the backdoor samples by modifying the node features in node classification and modifying the graph topology in graph classification.

GNN black-box watermarking methods present strong robustness against model post-processing techniques (\textbf{R3}) and offer advantages in overhead (\textbf{R4}). However, the integration of the backdoor introduces significant security vulnerabilities (\textbf{R1}). Furthermore, these methods only possess zero-bit capacity (\textbf{R2}), which requires that every watermarked model be associated with a unique trigger graph.
Ownership verification requires multiple queries to the suspect model for every trigger graph, which increases financial cost and is prone to detection. In addition, the watermark built on the backdoor is easily forged because the adversary can collect many naturally misclassified samples and falsely claim ownership~\cite{shao2024explanation}.

\noindent \textbf{GNN Fingerprinting.} GNN fingerprinting verifies ownership by comparing the similarity of inherent characteristics (i.e., fingerprints). Fingerprints can be layer-wise embeddings of nodes~\cite{waheed2024grove} or overall model behaviors~\cite{zhou2024revisiting}. 
Comparing similarity does not require embedding a backdoor (\textbf{R1}).
However, constructing fingerprints incurs a significant computational overhead (\textbf{R4}). Independently trained models may exhibit similarities to the fingerprinted model due to post-processing techniques (e.g., pruning, fine-tuning), leading to false positives (\textbf{R3}).
This method also possess zero-bit capacity (\textbf{R2}).

\section{Key Insights: LDDE}
\label{Insight}

\begin{figure}[!t]
    \centering
    \begin{subfigure}[t]{0.48\textwidth}
        \includegraphics[page=1, width=0.49\linewidth]{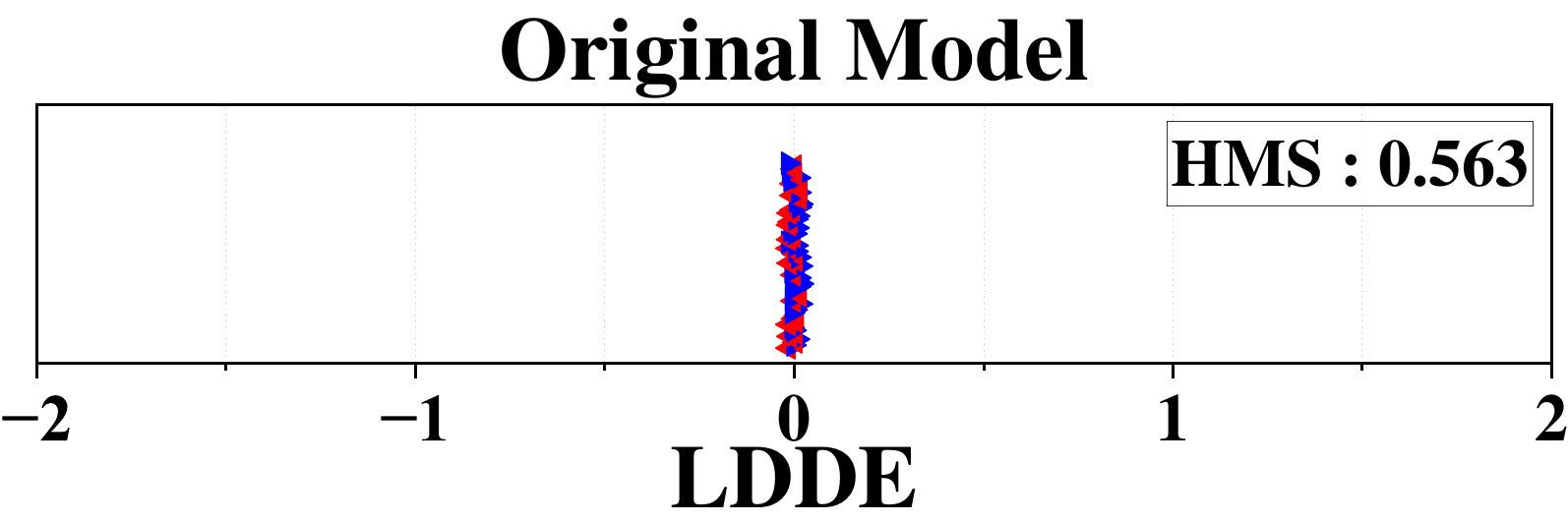}
        \includegraphics[page=2, width=0.49\linewidth]{fig/Insight2.pdf}
        \caption{Inductive paradigm}
    \end{subfigure}

    \begin{subfigure}[t]{0.48\textwidth}
        \includegraphics[page=5, width=0.49\linewidth]{fig/Insight2.pdf}
        \includegraphics[page=6, width=0.49\linewidth]{fig/Insight2.pdf}
        \caption{Transductive paradigm}
    \end{subfigure}
    
    \caption{Projections of LDDE values of selected edges before and after watermark embedding for GCNv2 trained on {\tt Physics} in \textit{S.T.A.}. Blue points represent edges targeted for positive LDDE signs, while red points represent those targeted for negative signs.}
    \label{fig:insight2}
    \vspace{-2mm}
\end{figure}

LDDE rises from the aggregation function ($AGG$) and the message exchange mechanisms ($MSG$) in GNNs. The LDDE indicates that, in a black-box query, a GNN model not only outputs the predictions of individual samples but also reveals the relational dependencies among them. This allows ownership verification to depend on the predictions of the samples or the relationships between them. We begin by defining LDDE and then present three critical observations. These observations form our insight into the suitability of LDDE as a watermark carrier. LDDE is defined as follows.

\begin{definition}
For two nodes $u$ and $v$ connected by an edge $e_{uv}$, the features of $u$ and $v$ are defined as $\mathbf{x}_u$ and $\mathbf{x}_v$, and the predictions of $u$ and $v$ are defined as $\mathbf{\hat{y}}_u$ and $\mathbf{\hat{y}}_v$. The difference between their node feature distance $\mathcal{D}(\mathbf{x}_u, \mathbf{x}_v) $ and their node prediction distance $\mathcal{D}(\mathbf{\hat{y}}_u,\mathbf{\hat{y}}_v)$ is LDDE. Formally, LDDE is defined as,
\begin{equation}
    \label{ldde}
    {\mathcal{F}_{L\!D\!D\!E}}(e_{uv}) = \mathcal{D}(\mathbf{\hat{y}}_u,\mathbf{\hat{y}}_v) - \mathcal{D}(\mathbf{x}_u, \mathbf{x}_v),
\end{equation}
\end{definition}

\noindent We use cosine similarity for $\mathcal{D}$ in this work as it is bounded. 

\noindent \textbf{Observation-1.} \textit{For the graph with $n$ edges inputted to a GNN model, $n$ LDDE values can be computed for the $n$ edges.}
GNNs leverage neighborhood aggregation and message-passing mechanisms to capture structural information from graph data~\cite{dai2024comprehensive}. These operations jointly leverage node features and graph topology, resulting in a predefined and measurable distance between connected nodes. Consequently, the LDDE value can be computed for every edge in any input graph to a GNN model. 

\noindent \textbf{Observation-2.} \textit{LDDE values of certain edges can be modified to align their signs with predefined positive or negative signs.} 
To embed the watermark string, the owner selects the edges that exhibit LDDE values close to zero and possess rare local topologies. The owner then fine-tunes the original model to jointly optimize both the primary task and the watermark embedding task. The latter modifies the LDDE values of the selected edges to match the designed watermark. 

We measure the Hamming similarity (HMS) between the extracted bit string and the target watermark. An HMS of 0.5 means random guessing, while an HMS of 1.0 means perfect match.  
Figure~\ref{fig:insight2} shows the variation of HMS for GCNv2 trained on {\tt Physics}. The LDDE values of the selected edges for the original models center on zero, with HMS falling in the range of 0.4 to 0.6. These LDDE values propose clearly positive or negative values in the watermarked model, with HMS achieving 1.0. These results show that the LDDE values can be modified to encode the intended watermark. 

\begin{figure}[!t]
    \centering
    \begin{subfigure}[t]{0.46\textwidth}
        \includegraphics[page=1, width=0.49\linewidth]{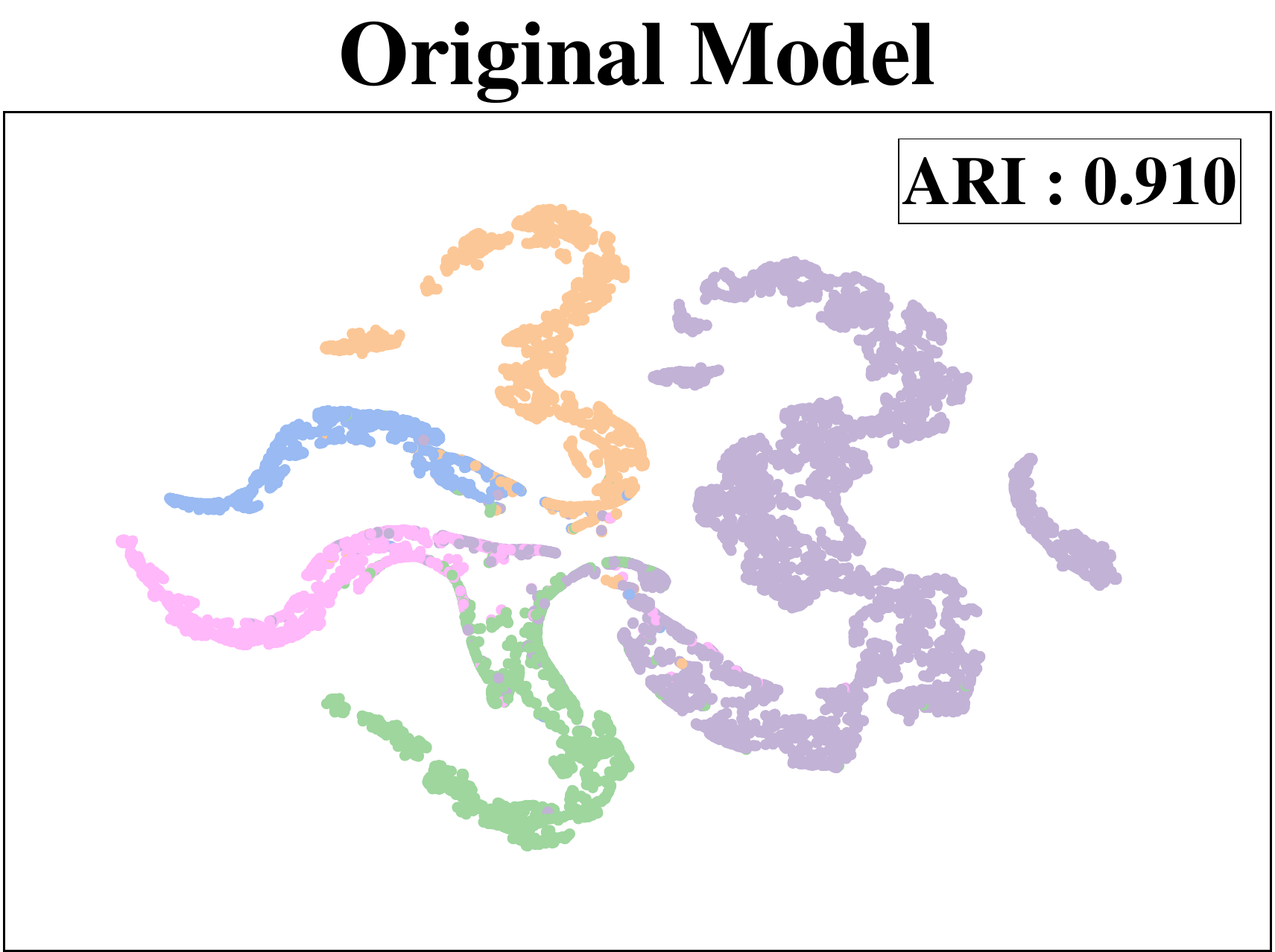}
        \includegraphics[page=2, width=0.49\linewidth]{fig/Insight3.pdf}
        \caption{Inductive paradigm}
    \end{subfigure}

    \begin{subfigure}[t]{0.46\textwidth}
        \includegraphics[page=5, width=0.49\linewidth]{fig/Insight3.pdf}
        \includegraphics[page=6, width=0.49\linewidth]{fig/Insight3.pdf}
        \caption{Transductive paradigm}
    \end{subfigure}
    
    \caption{The t-SNE projections of the predictions from both the original and watermarked models for GCNv2 on {\tt Physics} in \textit{S.T.A.}. Different colors indicate different classes.}
    \label{fig:insight3}
    \vspace{-2mm}
\end{figure}

\vspace{0.5mm}
\noindent \textbf{Observation-3.} \textit{Modifying the LDDE values of selected edges introduces a minor impact on the primary task.} In node-level GNNs, the distances between node pairs have been shown to correlate with the connectivity relationship: connected node pairs generally exhibit shorter distances than unconnected~\cite{he2021stealing, wu2022linkteller}. 
As this relationship serves as \textit{auxiliary information} that is not directly influenced by the primary task, it allows for minor adjustments without noticeably degrading the model's performance on the primary task.

Figure~\ref{fig:insight3} shows the impact of the LDDE modification on the primary task. We calculate the Adjusted Rand Index (ARI) \cite{steinley2004properties} of the test graph. An ARI of 1.0 means perfect classification. As shown in Figure~\ref{fig:insight3}, ARI drops from 0.910 to 0.883 in inductive and 0.938 to 0.899 in transductive. These results confirm that LDDE modifications have a minor impact on the model's primary task.

\section{\toolname: Design and implementation}
\label{Method}
\begin{figure*}[!ht]
    \centering
    \includegraphics[width=0.97\linewidth]{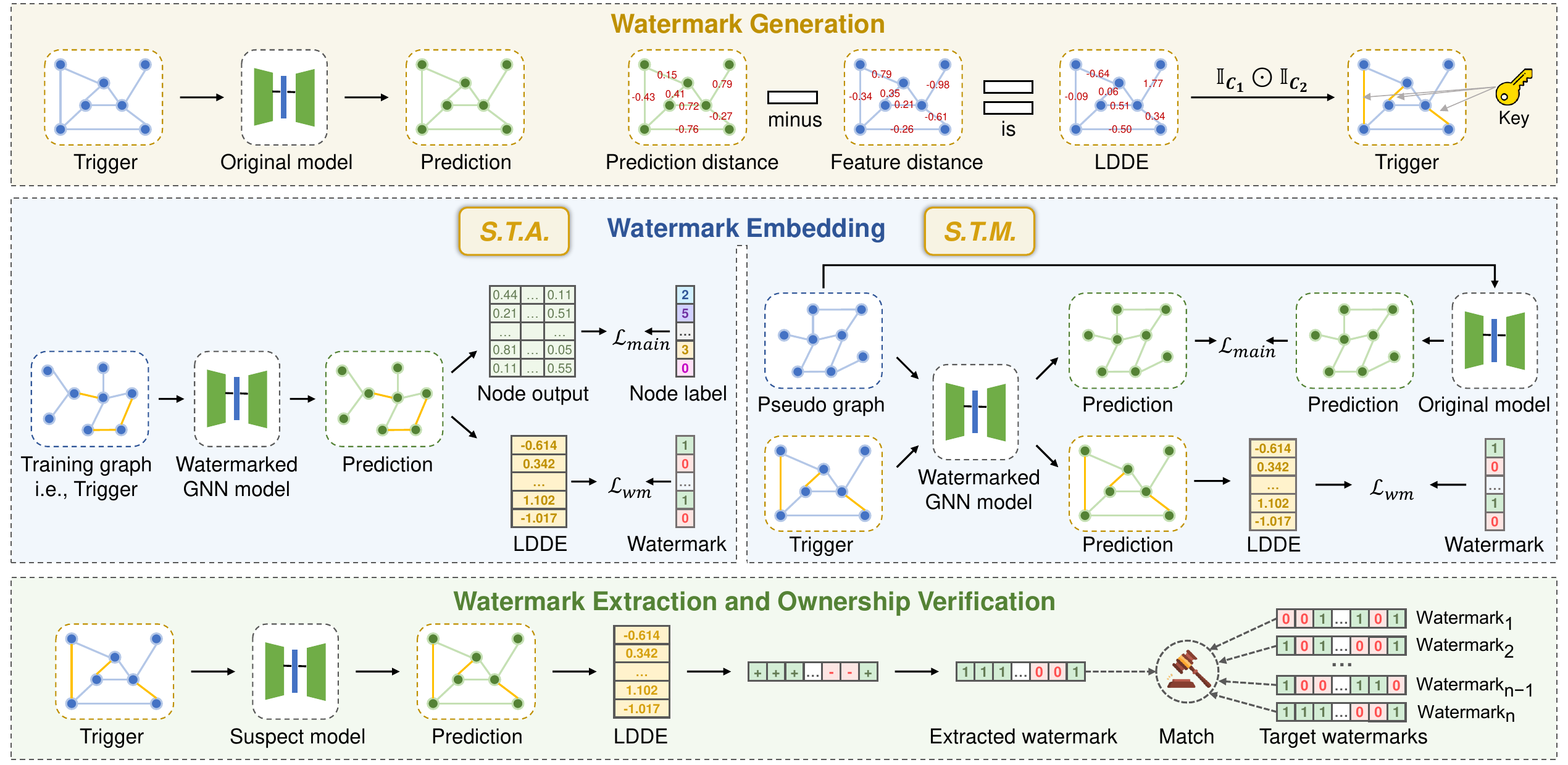}
    \caption{The overview of \toolname. The owner firstly selects the edges that satisfy \ref{cond1} and \ref{cond2} and marks them as the watermark key. The owner takes the training graph \textit{(S.T.A.)} or generates a graph \textit{(S.T.M.)} as the trigger graph. Both \textit{S.T.A.} and \textit{S.T.M.} share the same verification process.}
    \label{fig:Overall}
    \vspace{-3mm}
\end{figure*}

We first discuss other potential implementation methods for multi-bit or backdoor-free. 
We then describe the generation, embedding, and extraction of the watermark of \toolname.

\subsection{Starwman Solutions}
\label{subsec:starw}

One seemingly straightforward but ultimately inappropriate solution for GNN ownership verification is the direct adoption of black-box watermarking methods in computer vision models or natural language models~\cite{adi2018turning, Szyller2021}. For example, UBW~\cite{li2022untargeted} uses a particular value of entropy as a watermark. Trigger samples output correct classifications while exhibiting unusually high entropy. However, UBW substantially alters the sample output, severely degrading fidelity and increasing the likelihood of ownership misidentification for GNNs. EaaW~\cite{shao2024explanation} proposes a multi-bit black-box watermarking method by transforming the explanation of a specific
trigger sample into the watermark. However, due to the non-Euclidean nature of graph data, EaaW struggles with issues of fidelity and effectiveness. We compare EaaW with our method and show the results in Section \ref{Experiments}.
Thus, we explore GNN-specific properties for GNN black-box watermarking. 

Previous studies~\cite{he2021stealing, wu2022linkteller} show that the distance between a pair of nodes can reveal the existence of an edge. With the same GNN model, a pair of nodes with or without an edge yields different output distances. Inspired by this observation, we fix the input graph and make different watermarked GNN models produce distinct node output distances. The unique node output distances serve as the watermark to identify the watermarked GNN model. 
We map positive values of LDDE to ``1" and negative values to ``0". 
To ensure a balanced distribution of positive and negative values, we define and use LDDE rather than directly using node output distances.

\subsection{Watermark Generation}
We begin by delineating the watermark generation, including the method of generating the watermark bit string, and the rules of selecting proper edges as watermark carriers.

\noindent \textbf{Generate a multi-bit string as the watermark.} The watermark acts as a unique identifier for GNNs, with its multi-bit sequence constituting an identification number. In this work, we generate an $N_w$-bit watermark string by independently sampling each bit from a Bernoulli distribution with equal probability for 0 and 1 (i.e., Bernoulli(0.5)),
\begin{equation}
    \mathbf{w} = (w_1, w_2, \dots, w_{N_w}), \quad w_i \sim \text{Bernoulli}(0.5)
\end{equation}
where $\mathbf{w}$ is the watermark string; $N_w$ means the length of $\mathbf{w}$; $w_i$ is a bit of $\mathbf{w}$ which is either 0 or 1; and Bernoulli(0.5) is Bernoulli distribution equal probability.
In fact, the watermark string can encode arbitrary information.

\vspace{1mm}
\noindent \textbf{Transfer LDDE values to a binary string.}
For mapping rules, two fundamental problems must be addressed. The first problem is selecting the proper edges to carry watermarks. 
The second problem is normalizing the predictions.
 
For the first problem, although every edge possesses an LDDE value, not all edges are suitable as watermark carriers. 
If an edge has a prevalent local topology, modifying its LDDE affects all edges with similar structures. If an edge possesses a large LDDE value, altering its sign (from positive to negative or vise versa) requires substantial model adjustments.
Thus, edges suitable for carrying the watermark should satisfy the following conditions.

\begin{enumerate}[label=\textbf{C\arabic*}:, ref=\textbf{C\arabic*}, leftmargin=*]
\item \textbf{\label{cond1}} \textit{The edges possess statistically rare local structures.}
\item \textbf{\label{cond2}} \textit{The edges possess LDDE values close to zero.}
\end{enumerate}

Edges that better satisfy condition~\ref{cond1} minimize the impact of modifying LDDE values on other edges. Edges that better satisfy condition~\ref{cond2} require smaller adjustments to alter the signs of their LDDE values. 
We use the watermark key $\mathbf{k}$ to indicate the selected edges in the trigger graph.

For the second problem, the suspect model deployed in the cloud typically returns softmax-normalized posterior probabilities~\cite{sun2023deep}. 
The output in such a form exhibits minimal variation across dimensions and is always positive. The cosine similarity calculated on this form is close to zero for all node pairs.
We applied a scaling transformation to the posterior probabilities before calculating LDDE to expand the differences~\cite{luo2021feature}.
Specifically, we applied the following scaling transformation to $\mathbf{\hat{y}}$ in Equation~(\ref{ldde}):
\[    \mathbf{\tilde{y}} = \frac{\log(\mathbf{\hat{y}}) - \mu(\log(\mathbf{\hat{y}}))}{\sigma(\log(\mathbf{\hat{y}}))}
\]
\noindent The $\mu$ and $\sigma$ denote the mean and standard deviation in dimensions, respectively. For example, for the node $u$ with features $\mathbf{x}_u=[1, 2, 3, 4, 5]$ and predictions $\mathbf{\hat{y}}_u=[0.2, 0.3, 0.5]$, the node $v$ with features $\mathbf{x}_v=[6, 7, 8, 9, 10]$ and prediction $\mathbf{\hat{y}}_v=[0.3, 0.3, 0.4]$, the LDDE of $e_{uv}$ is calculated as following:
\vspace{-2mm}
\begin{align*}
    &\mathbf{\tilde{y}}_u = \frac{\log(\mathbf{\hat{y}}_u) - \mu(\log(\mathbf{\hat{y}}_u))}{\sigma(\log(\mathbf{\hat{y}}_u))} \approx [-1.1753,-0.0937,1.2689] \\
    &\mathbf{\tilde{y}}_v = \frac{\log(\mathbf{\hat{y}}_v) - \mu(\log(\mathbf{\hat{y}}_v))}{\sigma(\log(\mathbf{\hat{y}}_v))} \approx [-0.7071,-0.7071,1.4142] \\
    &\mathcal{F}_{L\!D\!D\!E} = \mathcal{D}(\mathbf{\tilde{y}}_u,\mathbf{\tilde{y}}_v) - \mathcal{D}(\mathbf{x}_u, \mathbf{x}_v) \approx -0.0674 
\end{align*}
\noindent $\mathcal{D}$ is cosine similarity. Without this transformation, the LDDE between the nodes $u$ and $v$ is 0.0093. The transformation makes cosine similarity away from zero and could be negative.

\subsection{Watermark Embedding}
When the owner trains the original model on the training graph, the training graph can directly serve as the trigger graph \textit{(S.T.A. scenario)}.
However, ownership transfer is common in real-world applications. The model owner may submit the model to a platform for release. The platform becomes the new owner and possesses the original model without the training graph. In such cases, the new owner has to generate a graph as the trigger graph and embed watermarks without the training graph of the original model \textit{(S.T.M. scenario)}.

\noindent \textbf{Embed watermarks in the setting that the training graph is available} \textit{(S.T.A.)}:
Prior research indicates that most of the connected nodes tend to share the same labels~\cite{he2021stealing}. In other words, edge-connected nodes with different labels have rare local structures. When using the training graph $\mathbf{G}$ as the trigger graph $\mathbf{T}$, condition~\ref{cond1} can be satisfied by selecting edges that connect nodes with different labels. To satisfy condition~\ref{cond2}, the LDDE values of the edges that satisfy condition~\ref{cond1} are calculated firstly, and then the $N_w$ edges with the smallest absolute LDDE values are selected.
The selected edges are denoted as the watermark key $\mathbf{k}$, which is a subset of the edges of the trigger graph $\mathbf{k} \subseteq \mathbf{E_T}$.

As illustrated in Figure~\ref{fig:Overall}, the watermarked model $\mathcal{M}_w$ is initialized by the original model $\mathcal{M}_o$.
To embed the watermark $\mathbf{w}$, the owner fine-tunes the watermarked model $\mathcal{M}_w$ to optimize the following objective function until $\mathbf{w}$ is successfully extracted from $\mathcal{M}_w$:
\begin{equation}
\min_{\mathcal{M}_w} \mathcal{L} =  \mathcal{L}_{\text{main}}(\mathcal{M}_w(\mathbf{G}), \mathbf{Y}) +  \mathcal{L}_{\text{wm}}\left(\mathbf{\hat{w}}, \mathbf{w} \right)
\end{equation}
\noindent where $\mathbf{G}$ is the training graph and $\mathbf{Y}$ is the node labels. $\mathcal{L}_{\text{main}}$ represents the primary task, which is the cross-entropy loss in this work. $\mathcal{L}_{\text{wm}}$ measures the dissimilarity between the extracted watermark $\mathbf{\hat{w}}$ and the target watermark $\mathbf{w}$, which is the binary cross-entropy loss in this work. 
$\mathbf{\hat{w}}$ is a vector formed by the LDDE values of edges in $\mathbf{k}$, i.e.
$\mathbf{\hat{w}}=[\mathcal{F}_{L\!D\!D\!E}(e_i)]_{i=1}^n \ \text{with} \ \{e_1,e_2,...,e_n \in \mathbf{k}\}$.

\vspace{1mm}
\noindent \textbf{Embed watermarks in the setting that the training graph is missing} \textit{(S.T.M.)}:
If the original model has been transferred to another entity, the new owner has the original model only without the training graph.
Lacking the training graph, the owner has to generate a graph as the trigger graph.
The initial node features of the trigger graph are randomly sampled from a Gaussian distribution, with feature dimensions matching the model input channels. The topology can be generated randomly or from a real graph for better concealment.
The node features are then updated for the following objective:
\begin{equation}
    \min_{\mathbf{X_T}} \mathcal{L}=  ||\mathbf{v}|| + \lambda_1 \cdot \left\|\frac{1}{1-\mathcal{D}(\mathbf{x}_u,\mathbf{x}_v)}\right\|
    \label{eq:stm}
\end{equation}

\noindent $\mathbf{v}$ denotes all LDDE values of the trigger graph $\mathbf{T}$, which is $\mathbf{v}=[\mathcal{F}_{L\!D\!D\!E}(e_i)]_{i=1}^n \ \text{with} \ \{e_1,e_2,...,e_n \in \mathbf{E_T}\}$.
$\mathcal{D}(\mathbf{x}_u,\mathbf{x}_v)$ denotes all cosine similarity between every connected node pairs. $||\!\cdot\!||$ is the L1 norm.
The first term forces all LDDE values close to zero for the condition~\ref{cond2}.
The second term imposes a penalty to prevent node feature distances from approaching the boundaries of the cosine similarity range $[-1,+1]$. When the node feature distance nears the cosine similarity boundary, it becomes challenging to flip the signs of the LDDE values, even they are close to zero. $\lambda_1$ is a regularization coefficient.
Once the update completes, we applied Node2Vec~\cite{grover2016node2vec} and DBSCAN~\cite{schubert2017dbscan} to identify edges that meet the condition~\ref{cond1} due to the absence of labels. Subsequently, we select $N_w$ edges that meet condition~\ref{cond2} and mark them as $\mathbf{k}$. $\mathbf{k}$ is a subset of the edges of the trigger graph $\mathbf{k} \subseteq \mathbf{E_T}$.

\begin{algorithm}[!t]
    \caption{Ownership Verification}
    \label{alg:smv}
    \SetKwInOut{Input}{\textbf{Input}}
    \SetKwInOut{Output}{\textbf{Output}}

    \Input{
        API of the suspect model $\mathcal{M}_s$; 
        Trigger graph $\mathbf{T}$; 
        Watermark Key $\mathbf{k}$;
        Watermark set $\mathbf{S}_w = \{\mathbf{w}_i \}_{i=1}^n$;
    }
    \Output{
        (1) A boolean value: True if $\mathcal{M}_s$ is identified as an unauthorized copy; (2) The matched watermark string, returned only if True.
    }
    \BlankLine

    $\mathbf{\hat{Y}} \leftarrow \mathcal{M}_s(\mathbf{T})$ \label{alg2line1}\;
    $\mathbf{\hat{w}}=[\mathcal{F}_{L\!D\!D\!E}(e_i)]_{i=1}^n \ \text{with} \ \{e_1,e_2,...,e_n \in \mathbf{k}\}$ \;
    $\mathbf{\tilde{w}} \gets Signal(\mathbf{\hat{w}}) $ \label{alg2line3}\;

    $\mathbf{w}_m \leftarrow \arg\max \text{SIM}(\mathbf{\tilde{w}}, \mathbf{S}_w)$ \;
    \tcp{$\mathbf{w}_m \in \mathbf{S}_w$ is the most similar watermark to $\mathbf{\tilde{w}}$}
    
    \lIf{$\text{SIM}(\mathbf{w}_m, \mathbf{\tilde{w}}) \geq \tau$}{\Return True, $\mathbf{w}_m$}
    \lElse{\Return False}
\end{algorithm}

In the absence of the training graph $\mathbf{G}$, the challenge during watermark embedding lies in maintaining the performance of $\mathcal{M}_w$ in the primary task. To address this, we adapt the concept of data-free adversarial distillation (DFAD) for GNNs~\cite{deng2021graph,zhuang2022data}. 
Specifically, we adopt the Barabasi-Albert (BA) model~\cite{dehmamy2019understanding} to randomly generate the pseudo graph topology $\mathbf{E_{\tilde{G}}}$, and initialize the node features $\mathbf{X_{\tilde{G}}}$ with samples drawn from a Gaussian distribution. $\mathbf{\tilde{G}}$ acts as a substitute for $\mathbf{G}$. During watermark embedding, after each update to $\mathcal{M}_w$, $\mathbf{X_{\tilde{G}}}$ is updated to maximize the output difference between the original model $\mathcal{M}_o$ and the watermarked model $\mathcal{M}_w$. The optimization objective is:
\begin{equation}
    \min_{\mathcal{M}_w} \max_{\mathbf{X_{\tilde{G}}}}  \mathcal{L}=  \mathcal{L}_{\text{main}}(\mathcal{M}_w(\mathbf{\tilde{G}}), \mathcal{M}_o(\mathbf{\tilde{G}})) +   \mathcal{L}_{\text{wm}}(\mathbf{\hat{w}},\mathbf{w})
\end{equation}

\noindent where $\mathcal{M}_o$ denotes the original model and $\mathcal{M}_w$ denotes the watermarked model. The pseudo graph  
$\mathbf{\tilde{G}}$ serves as a substitute for the training graph $\mathbf{G}$ to maintain the primary task performance of the watermarked model. $\mathcal{L}_{\text{main}}$ aims to minimize the prediction discrepancy between the watermarked model and the original model on the pseudo graph $\mathbf{\tilde{G}}$. $\mathcal{L}_{\text{wm}}$ is used to embed the watermark. The node features of the pseudo graph $\mathbf{X_{\tilde{G}}}$ are updated during training, increasing the learning challenge for the watermarked model and encouraging it to learn more robust features from the original model.

Unlike classical DFAD for GNNs~\cite{zhuang2022data}, which updates both the node features and the graph topology, \textit{S.T.M.} keeps the graph topology fixed and updates only the node features. This is because LDDE is relevant to the graph topology. Altering the topology would increase the uncertainty of modifying the LDDE values.
The principle of \textit{S.T.M.} dictates that the optimal points of the watermarked model and the original model are nearly identical, allowing the original model to be transformed into the watermarked model with only minor alteration.

\subsection{Watermark Extraction and Ownership Verification:}
Both \textit{S.T.A.} and \textit{S.T.M.} have the same process of watermark extraction and ownership verification.
As shown in Figure~\ref{fig:Overall}, the verifier queries the suspect model’s API using the trigger graph to obtain the predictions for all nodes. The verifier then computes the LDDE values of the edges in the watermark key to form a sequence. This sequence is binarized to recover the embedded watermark. The complete extraction procedure is detailed in Algorithm~\ref{alg:smv}.

$Signal(x)$ returns 0 if $x < 0$ and 1 otherwise. SIM stands for similarity. $\tau$ is the judgment threshold and can be selected according to the normal distribution. It is noticed that if the suspect model is an unauthorized copy, the return is the original watermark in $\mathbf{S}_w$, not the extracted one. 
Section~\ref{Experiments} details the judgment threshold $\tau$.

\section{Experiments}
\label{Experiments}
We evaluate \toolname in six datasets and six mainstream models to address research questions (RQs) as follows.

\begin{itemize}[leftmargin=*]
    \item \textbf{RQ1.} The effectiveness and fidelity of \toolname. Can \toolname distinguish exactly watermarked models and independently trained models? How much accuracy decreases the watermarked model compared to the original model on the primary tasks? 
    \item \textbf{RQ2.} The robustness of \toolname. We involve 4 ownership attacks such as pruning, fine-tuning, watermark overwriting, and model extraction.
    \item \textbf{RQ3.} The impact of hyper-parameters on \toolname. We investigate two key hyper-parameters: the watermark length $N_w$ and similar threshold $\tau$.
\end{itemize}

We first describe the experimental setting and then demonstrate the experimental results.

\subsection{Experiments Setting}
\noindent \textbf{Environment Configurations.}
We conduct the experiments on an NVIDIA A100 GPU with 80GB RAM. We used PyTorch 2.5.1 and PyG 2.6.1 as the programming languages.

\noindent \textbf{Dataset.} We conduct experiments on six datasets as shown in Table \ref{tab:dataset}. Each dataset is represented as a graph, where each node corresponds to a sample. For each graph, i.e., a dataset, referring to previous implementations~\cite{waheed2024grove, zhou2024revisiting}, we split 40\% of the nodes for the training graph, 30\% of the nodes for the adversary graph held by the adversary, and 30\% of the nodes for the test graph.

\noindent \textbf{Model Architectures.}
We use different model architectures for different datasets. Specially, we use GAT~\cite{velickovic2017graph} for {\tt Cora}, Graph Transformer (GTF)~\cite{shi2020masked} for {\tt DBLP}, GraphSAGE (SAGE)~\cite{hamilton2017inductive} for {\tt Photo}, SSG~\cite{zhu2021simple} for {\tt CS}, GCNv2~\cite{chen2020simple} for {\tt Physics}, and ARMA\cite{bianchi2021graph} for {\tt Blog}. Except for GCNv2, each architecture includes four graph convolution layers, with the final layer serving as the output layer. GCNv2 is composed of an input linear layer, three graph convolution layers, and an output linear layer. All models use the ELU activation function, with the Adam optimizer set to a learning rate of 1e-4 and a weight decay of 1e-4. Each model is trained for 500 iterations, and we use the cross-entropy loss as the objective for the primary tasks. The $\lambda_1$ in Eq. (\ref{eq:stm}) is 1e-4. 

\noindent \textbf{Metrics.}
We use the following metrics to evaluate \toolname:
\begin{itemize}[leftmargin=*]
    \item \textit{Test Accuracy (TAC):} The accuracy of the model on the test graph, representing the GNN performance on primary tasks.
    \item \textit{Hamming similarity (HMS):} Hamming similarity quantifies the similarity between two binary strings of equal length. Formally, given two binary sequences $\mathbf{\tilde{w}},\mathbf{w} \in \{0,1\}^{N_w}$, the HMS is defined as:
    \begin{equation}
        \label{equ:HMS}
        \text{HMS}(\mathbf{\tilde{w}},\mathbf{w}) = \frac{1}{N_w} \sum^{N_w}_{i=1} \mathbb{I}\{\mathbf{\tilde{w}}[i] = \mathbf{w}[i]\}
    \end{equation}
    where $\mathbb{I}(x)$ is an indicator function that returns 1 if $x$ is true and 0 otherwise. $\mathbf{\tilde{w}}$ is the extracted watermark string, while $\mathbf{w}$ is the target watermark string. $N_w$ is the length of the watermark string. HMS ranges from 0 to 1, with 1 indicating perfect agreement and 0.5 meaning random guessing. HMS with 0 means that every bit is opposite to the target watermark.
    \item \textit{Ownership Verification Accuracy (OVA):} Correct detection rate of ownership verification.
    \item \textit{False Positive Rate (FPR):} FPR reports the rate of independently trained models mistakenly identified as unauthorized copies. It must remain zero; any value above zero indicates that the ownership verification audit is untrustworthy.
\end{itemize}

For the fingerprinting method (i.e. RBOVG), the ownership verification results are given by the classifier. For the watermarking methods (i.e. WGB, EaaW and \toolname), the ownership verification results depend on similar threshold.
We adopted the midpoint of each similarity metric as the similar threshold, \(\tau\): a similarity score above \(\tau\) indicates an unauthorized copy. Specifically, $\tau$ is set to 0.5 for WGB (backdoor activation rate) and 0.75 for \toolname and EaaW (hamming similarity). 

\begin{table}[!t]
    \centering
    \caption{Datasets statistics of node classification task.}
    \label{tab:dataset}
    \setlength{\tabcolsep}{8pt}{
    \resizebox{0.47\textwidth}{!}{
    \begin{tabular}{lrrrc}  
        \toprule
        \textbf{Datasets} & \textbf{Nodes} & \textbf{Edges} & \textbf{Features} & \textbf{Classes} \\
        \midrule
        {\tt Cora}~\cite{yang2016revisiting} & 2,708 & 10,556 & 1,433 & 7 \\
        {\tt DBLP}~\cite{bojchevski2017deep} & 17,716 & 105,734 & 1,639 & 4 \\
        {\tt Photo}~\cite{shchur2018pitfalls} & 7,650 & 238,162 & 745 & 8\\
        {\tt CS}~\cite{shchur2018pitfalls} & 18,333 & 163,788 & 6,805 & 15 \\
        {\tt Physics}~\cite{shchur2018pitfalls} & 34,493 & 495,924 & 8,415 & 5 \\
        {\tt Blog}~\cite{yang2020scaling} & 5,196 & 343,486 & 8,189 & 6 \\
        \bottomrule
    \end{tabular}
    }}
\end{table}

\noindent \textbf{Baseline.} We select WGB~\cite{xu2023watermarking} and RBOVG~\cite{zhou2024revisiting} as baselines. These methods are the state-of-the-art in GNN black-box watermarking and fingerprinting, respectively. We also test whether EaaW~\cite{shao2024explanation} mentioned in Section~\ref{subsec:starw} can be directly transplanted to GNN. 

\begin{table*}[!ht]
\centering
\caption{Effectiveness of \toolname and baselines. We report ownership verification accuracy (OVA) (\%) and false positive rate (FPR) (\%). For watermarking methods, we additionally report the first and third quartiles of the similarity metrics (Q1/3). The similarity metric is backdoor activation rate (BAR) for WGB and Hamming similarity (HMS) for EaaW and \toolname.}
\label{tab:effectiveness}
\setlength{\tabcolsep}{3pt}
{
\resizebox{1.0\textwidth}{!}{
\begin{NiceTabular}{c|lcccccccccccccccccc}
\toprule
                              & \multirow{2}{*}{\textbf{Datasets}} & \multicolumn{2}{c}{\textbf{RBOVG}} & \multicolumn{4}{c}{\textbf{WGB} {\tt (BAR $\tau=0.5$)}}                 & \multicolumn{4}{c}{\textbf{EaaW}{\tt (HMS $\tau=0.75$)}}                & \multicolumn{4}{c}{\textbf{\toolname} (\textit{S.T.A.}){\tt (HMS $\tau=0.75$)}}            & \multicolumn{4}{c}{\textbf{\toolname} (\textit{S.T.M.}) {\tt (HMS $\tau=0.75$)}}            \\
\cmidrule(lr){3-4} \cmidrule(lr){5-8} \cmidrule(lr){9-12} \cmidrule(lr){13-16} \cmidrule(lr){17-20}
                              &                           & OVA         & FPR         & OVA  & FPR  & {\!\!$\mathcal{M}_i$(Q1/3)\!\!} & {\!\!$\mathcal{M}_w$(Q1/3)\!\!} & OVA  & FPR  & {\!\!$\mathcal{M}_i$(Q1/3)\!\!} & {\!\!$\mathcal{M}_w$(Q1/3)\!\!} & OVA  & FPR  & {\!\!$\mathcal{M}_i$(Q1/3)\!\!} & {\!\!$\mathcal{M}_w$(Q1/3)\!\!} & OVA  & FPR  & {\!\!$\mathcal{M}_i$(Q1/3)\!\!} & {\!\!$\mathcal{M}_w$(Q1/3)\!\!} \\
\midrule
\multirow{6}{*}{\rotatebox{-90}{\textbf{Inductive}}}    & {\tt Cora}                      & 100.0      & 0.0      & 100.0 & 0.0 & 0.00/0.01   & 0.79/0.89   & 84.0  & 0.0  & 0.48/0.58   & 0.73/0.84   & 100.0 & 0.0 & 0.46/0.55   & 0.92/0.98   & 100.0 & 0.0 & 0.47/0.56   & 1.00/1.00   \\
                              &{\tt DBLP}                      & 100.0      & 0.0      & 100.0 & 0.0 & 0.00/0.00   & 1.00/1.00   & 51.0  & 0.0  & 0.44/0.54   & 0.52/0.63   & 100.0 & 0.0 & 0.47/0.55   & 1.00/1.00   & 100.0 & 0.0 & 0.45/0.55   & 1.00/1.00   \\
                              & {\tt Photo}                     & 100.0      & 0.0      & 100.0 & 0.0 & 0.00/0.01   & 0.99/1.00   & 66.0  & 0.0  & 0.47/0.55   & 0.70/0.77   & 100.0 & 0.0 & 0.50/0.54   & 0.88/0.93   & 100.0 & 0.0 & 0.43/0.55   & 1.00/1.00   \\
                              & {\tt CS}                        & 100.0      & 0.0      & 100.0 & 0.0 & 0.00/0.01   & 1.00/1.00   & 91.0  & 0.0  & 0.47/0.54   & 0.77/0.93   & 100.0 & 0.0 & 0.47/0.53   & 0.98/0.98   & 100.0 & 0.0 & 0.47/0.55   & 1.00/1.00   \\
                              & {\tt Physics}                   & 100.0      & 0.0      & 100.0 & 0.0 & 0.00/0.01   & 1.00/1.00   & 100.0 & 0.0  & 0.45/0.60   & 0.98/0.99   & 100.0 & 0.0 & 0.47/0.55   & 0.98/1.00   & 100.0 & 0.0 & 0.48/0.55   & 1.00/1.00   \\
                              & {\tt Blog}                      & 100.0      & 0.0      & 100.0 & 0.0 & 0.08/0.11   & 1.00/1.00   & 100.0 & 0.0  & 0.56/0.61   & 0.99/1.00   & 100.0 & 0.0 & 0.45/0.56   & 1.00/1.00   & 100.0 & 0.0 & 0.47/0.57   & 1.00/1.00   \\
\midrule
\midrule
\multirow{6}{*}{\rotatebox{-90}{\textbf{Transductive}}} & {\tt Cora}                      & 100.0      & 0.0      & 100.0 & 0.0 & 0.00/0.02   & 0.93/0.94   & 100.0 & 0.0  & 0.48/0.53   & 0.89/0.95   & 100.0 & 0.0 & 0.46/0.53   & 0.95/0.97   & 100.0 & 0.0 & 0.45/0.53   & 0.97/1.00   \\
                              & {\tt DBLP}                      & 100.0      & 0.0      & 100.0 & 0.0 & 0.01/0.12   & 1.00/1.00   & 73.0  & 0.0  & 0.45/0.55   & 0.71/0.83   & 100.0 & 0.0 & 0.47/0.55   & 1.00/1.00   & 100.0 & 0.0 & 0.45/0.53   & 0.97/0.98   \\
                              & {\tt Photo}                     & 100.0      & 0.0      & 100.0 & 0.0 & 0.00/0.01   & 0.96/1.00   & 100.0 & 0.0  & 0.50/0.59   & 0.98/1.00   & 100.0 & 0.0 & 0.45/0.52   & 0.95/0.98   & 100.0 & 0.0 & 0.44/0.56   & 1.00/1.00   \\
                              & {\tt CS}                        & 100.0      & 0.0      & 100.0 & 0.0 & 0.00/0.01   & 1.00/1.00   & 100.0 & 0.0  & 0.61/0.66   & 0.99/1.00   & 100.0 & 0.0 & 0.47/0.55   & 1.00/1.00   & 100.0 & 0.0 & 0.45/0.53   & 1.00/1.00   \\
                              & {\tt Physics}                   & 100.0      & 0.0      & 100.0 & 0.0 & 0.00/0.01   & 1.00/1.00   & 95.0  & 10.0 & 0.58/0.69   & 1.00/1.00   & 100.0 & 0.0 & 0.45/0.53   & 1.00/1.00   & 100.0 & 0.0 & 0.45/0.56   & 1.00/1.00   \\
                              & {\tt Blog}                      & 100.0      & 0.0      & 100.0 & 0.0 & 0.04/0.11   & 1.00/1.00   & 100.0 & 0.0  & 0.54/0.59   & 1.00/1.00   & 100.0 & 0.0 & 0.51/0.55   & 1.00/1.00   & 100.0 & 0.0 & 0.45/0.52   & 1.00/1.00   \\
\bottomrule
\end{NiceTabular}
}
}
\end{table*}

\begin{table*}[!ht]
\centering
\caption{The test accuracy (TAC) (\%) of the original model ($\mathcal{M}_o$) and the watermarked (or fingerprinted) models. Values following $\pm$ denote the standard deviation (unbiased estimate). `$\downarrow \uparrow \mathcal{M}_o$' indicates the TAC change of the watermarked (or fingerprinted) models compared to the original model.}
\label{tab:fidelity}
\setlength{\tabcolsep}{8pt}{
\resizebox{1.0\textwidth}{!}{
\begin{NiceTabular}{l|lccccccccccc}
\toprule
        & {\multirow{2}{*}{\textbf{Datasets}}} & {\multirow{2}{*}{\textbf{$\mathcal{M}_o$}}} & \multicolumn{2}{c}{\textbf{WGB}} & \multicolumn{2}{c}{\textbf{RBOVG}} & \multicolumn{2}{c}{\textbf{EaaW}} & \multicolumn{2}{c}{\textbf{\toolname} \textit{(S.T.A.)}} & \multicolumn{2}{c}{\textbf{\toolname} \textit{(S.T.M.)}} \\
\cmidrule(lr){4-5}\cmidrule(lr){6-7}\cmidrule(lr){8-9}\cmidrule(lr){10-11}\cmidrule(lr){12-13}
          &                          &                       & TAC           & $\downarrow \uparrow \mathcal{M}_o$      & TAC             & $\downarrow \uparrow \mathcal{M}_o$      & TAC            & $\downarrow \uparrow \mathcal{M}_o$      & TAC              & $\downarrow \uparrow \mathcal{M}_o$        & TAC              & $\downarrow \uparrow \mathcal{M}_o$        \\
\midrule
\multirow{6}{*}{\rotatebox{-90}{\textbf{Inductive}}}    & {\tt Cora}                                          & 77.58                                     & 71.65{\tiny $\pm$1.72}   & \textcolor{red}{$\downarrow$5.93}    & 71.10{\tiny $\pm$2.25}     & \textcolor{red}{$\downarrow$6.48}    & 69.40{\tiny $\pm$1.80}    & \textcolor{red}{$\downarrow$8.18}    & 74.99{\tiny $\pm$0.82}      & \textcolor{red}{$\downarrow$2.59}      & 77.04{\tiny $\pm$0.56}      & \textcolor{red}{$\downarrow$0.54}      \\
                              & {\tt DBLP}                                          & 75.54                                     & 73.09{\tiny $\pm$0.42}   & \textcolor{red}{$\downarrow$2.45}    & 72.24{\tiny $\pm$0.40}     & \textcolor{red}{$\downarrow$3.31}    & 71.99{\tiny $\pm$0.36}    & \textcolor{red}{$\downarrow$3.55}    & 74.50{\tiny $\pm$0.32}      & \textcolor{red}{$\downarrow$1.04}      & 75.51{\tiny $\pm$0.14}      & \textcolor{red}{$\downarrow$0.03}      \\
                              & {\tt Photo}                                         & 92.51                                     & 92.07{\tiny $\pm$0.65}   & \textcolor{red}{$\downarrow$0.44}    & 90.23{\tiny $\pm$0.73}     & \textcolor{red}{$\downarrow$2.28}    & 89.68{\tiny $\pm$0.64}    & \textcolor{red}{$\downarrow$2.83}    & 89.30{\tiny $\pm$0.74}      & \textcolor{red}{$\downarrow$3.21}      & 90.76{\tiny $\pm$1.97}      & \textcolor{red}{$\downarrow$1.75}      \\
                              & {\tt CS}                                            & 93.67                                     & 93.81{\tiny $\pm$0.25}   & \textcolor{blue}{$\uparrow$0.14}   & 91.06{\tiny $\pm$1.54}     & \textcolor{red}{$\downarrow$2.61}    & 92.20{\tiny $\pm$0.40}    & \textcolor{red}{$\downarrow$1.47}    & 92.17{\tiny $\pm$0.35}      & \textcolor{red}{$\downarrow$1.50}      & 93.74{\tiny $\pm$0.10}      & \textcolor{blue}{$\uparrow$0.07}     \\
                              & {\tt Physics}                                       & 95.89                                     & 95.49{\tiny $\pm$0.07}   & \textcolor{red}{$\downarrow$0.40}    & 93.64{\tiny $\pm$0.33}     & \textcolor{red}{$\downarrow$2.25}    & 89.90{\tiny $\pm$0.26}    & \textcolor{red}{$\downarrow$5.99}    & 93.75{\tiny $\pm$0.34}      & \textcolor{red}{$\downarrow$2.14}      & 95.90{\tiny $\pm$0.05}      & \textcolor{blue}{$\uparrow$0.01}     \\
                              & {\tt Blog}                                          & 94.03                                     & 93.71{\tiny $\pm$1.67}   & \textcolor{red}{$\downarrow$0.32}    & 89.77{\tiny $\pm$0.52}     & \textcolor{red}{$\downarrow$4.26}    & 91.11{\tiny $\pm$0.53}    & \textcolor{red}{$\downarrow$2.92}    & 91.44{\tiny $\pm$0.65}      & \textcolor{red}{$\downarrow$2.59}      & 93.90{\tiny $\pm$0.26}      & \textcolor{red}{$\downarrow$0.13}      \\
\midrule
\midrule
\multirow{6}{*}{\rotatebox{-90}{\textbf{Transductive}}} & {\tt Cora}                                          & 82.14                                     & 80.62{\tiny $\pm$1.43}   & \textcolor{red}{$\downarrow$1.52}    & 81.57{\tiny $\pm$0.66}     &\textcolor{red}{$\downarrow$0.57}    & 73.40{\tiny $\pm$1.50}    &\textcolor{red}{$\downarrow$8.74}    & 78.67{\tiny $\pm$1.01}      & \textcolor{red}{$\downarrow$3.47}      & 81.22{\tiny $\pm$0.37}      & \textcolor{red}{$\downarrow$0.92}      \\
                              & {\tt DBLP}                                          & 83.97                                     & 81.56{\tiny $\pm$0.82}   & \textcolor{red}{$\downarrow$2.41}    & 79.59{\tiny $\pm$0.50}     & \textcolor{red}{$\downarrow$4.38}    & 77.10{\tiny $\pm$0.24}    & \textcolor{red}{$\downarrow$6.87}    & 81.68{\tiny $\pm$0.41}      & \textcolor{red}{$\downarrow$2.29}      & 83.65{\tiny $\pm$0.19}      & \textcolor{red}{$\downarrow$0.32}      \\
                              & {\tt Photo}                                         & 95.55                                     & 95.12{\tiny $\pm$0.44}   & \textcolor{red}{$\downarrow$0.43}    & 93.68{\tiny $\pm$0.56}     & \textcolor{red}{$\downarrow$1.87}    & 93.41{\tiny $\pm$0.33}    & \textcolor{red}{$\downarrow$2.14}    & 93.15{\tiny $\pm$0.67}      & \textcolor{red}{$\downarrow$2.40}      & 93.92{\tiny $\pm$1.22}      & \textcolor{red}{$\downarrow$1.63}      \\
                              & {\tt CS}                                            & 95.75                                     & 95.33{\tiny $\pm$0.20}   & \textcolor{red}{$\downarrow$0.42}    & 92.71{\tiny $\pm$0.87}     & \textcolor{red}{$\downarrow$3.04}    & 91.20{\tiny $\pm$0.60}    & \textcolor{red}{$\downarrow$4.55}    & 94.22{\tiny $\pm$0.43}      & \textcolor{red}{$\downarrow$1.53}      & 95.62{\tiny $\pm$0.09}      & \textcolor{red}{$\downarrow$0.13}      \\
                              & {\tt Physics}                                       & 97.04                                     & 96.98{\tiny $\pm$0.06}   & \textcolor{red}{$\downarrow$0.06}    & 93.98{\tiny $\pm$0.21}     & \textcolor{red}{$\downarrow$3.06}    & 92.90{\tiny $\pm$0.15}    & \textcolor{red}{$\downarrow$4.14}    & 94.82{\tiny $\pm$0.53}      & \textcolor{red}{$\downarrow$2.22}      & 97.03{\tiny $\pm$0.08}      & \textcolor{red}{$\downarrow$0.01}      \\
                              & {\tt Blog}                                          & 96.21                                     & 95.86{\tiny $\pm$0.89}   & \textcolor{red}{$\downarrow$0.35}    & 92.67{\tiny $\pm$0.56}     & \textcolor{red}{$\downarrow$3.54}    & 90.11{\tiny $\pm$0.24}    & \textcolor{red}{$\downarrow$6.10}    & 92.65{\tiny $\pm$0.63}      & \textcolor{red}{$\downarrow$3.56}      & 96.12{\tiny $\pm$0.11}      & \textcolor{red}{$\downarrow$0.09}   \\
\bottomrule
\end{NiceTabular}
}
}
\end{table*}

\subsection{Effectiveness and Fidelity (RQ1)}
\noindent \textbf{Experiment Design.} 
We fine-tune the original model to embed the watermark using the Adam optimizer with a learning rate of 5e-5 and a weight decay of 1e-4. The length of the watermark string $N_w$ is 64. We generated 50 watermarked models and 50 independently trained models. For WGB, we mask 20\% of node features and assign the label ``2” to create backdoor samples. For RBOVG, we follow the authors’ default settings, training 60 shadow surrogate models and 60 independently trained models for the classifier. The training graph is evenly split: one half is used to train fingerprinted models and the other to train shadow models. Consequently, only half of the training graph is available for training the fingerprinted model. The other settings are consistent.

\noindent \textbf{Result Analysis.}
As shown in Table~\ref{tab:effectiveness}, \toolname is effective in distinguishing watermarked models from independently trained models, with 100\% OVA and 0\% FPR in all cases. EaaW fails to achieve perfect classification, with OVA scores below 100\% in most cases, and even exhibiting FPR values greater than zero for {\tt Physics}. Despite effective in image and text domains, EaaW is not well-suited for GNNs since it does not account for the impact of edges, which are equally crucial as node features for GNNs.

Table~\ref{tab:fidelity} presents the fidelity results. \toolname (\textit{S.T.M.}) achieves the best fidelity, with an average test accuracy (TAC) drop of 0.45\%. The high fidelity is because \toolname (\textit{S.T.M.}) generates a graph as the trigger graph for the watermark task, allowing the watermark and primacy tasks to operate on different graphs and thus minimizing mutual interference. 
\toolname (\textit{S.T.A.}) exhibits better fidelity than RBOVG (3.13\%) but worse than WGB (1.21\%), with an average TAC drop of 2.37\%. EaaW incurs a large average TAC drop of 4.79\%.

\noindent \textbf{Answers to RQ1:} \toolname achieves perfect accuracy (100\%) in ownership verification. Regarding fidelity, assessed through average accuracy degradation, the performance ranking is: \toolname \textit{S.T.M.} (0.45\%) $<$ WGB (1.21\%) $<$ \toolname \textit{S.T.A.} (2.37\%) $\approx$ RBOVG (3.13\%) $<$ EaaW (4.79\%).

\subsection{Robustness against Ownership attacks (RQ2)}

\noindent \textbf{Experiment Design.}
After obtaining a copy of the watermarked model, the adversary may deploy it directly as Machine Learning as a Service (MLaaS) or employ ownership attacks to disrupt the watermark in the copy model for evading ownership detection. We consider four such attacks: pruning, fine-tuning, watermark overwriting, and model extraction.
The specific experimental settings are as follows:

\begin{figure}[!t]
    \centering
    \includegraphics[page=1, width=0.49\linewidth]{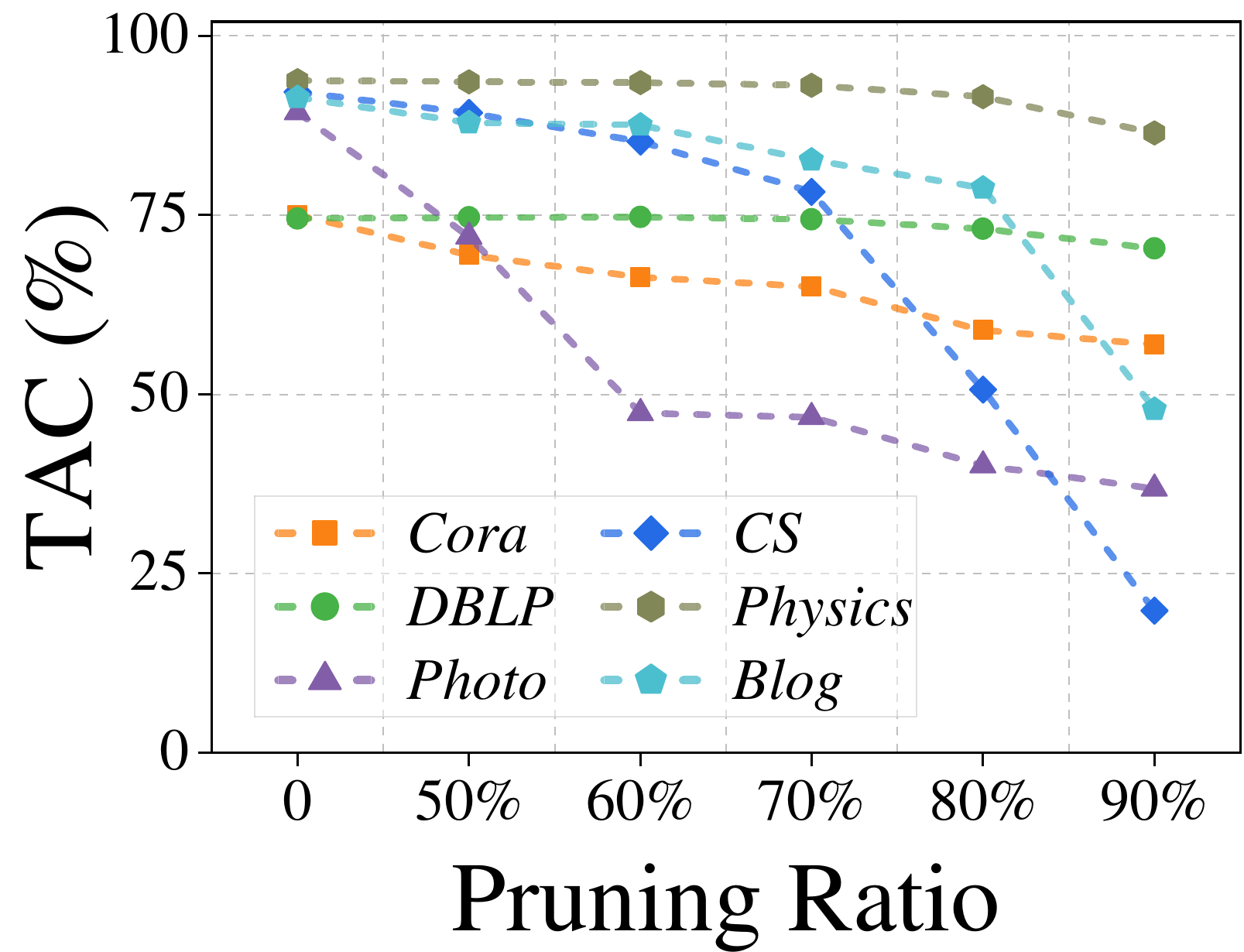}
    \includegraphics[page=2, width=0.49\linewidth]{fig/Pruning.pdf}
    \caption{Test accuracy (TAC) and Hamming similarity (HMS) with increasing pruning ratio. This figure depicts \toolname in \textit{S.T.A.} under the inductive paradigm.}
    \label{fig:pruning}
    \vspace{-2mm}
\end{figure}

\begin{table*}[ht]
\centering
\caption{Ownership verification accuracy (OVA) and false positive rate (FPR) after pruning both the independently trained models and the watermarked (or fingerprinted) models with 70\% ratio. We additionally report the first and third quartiles of the similarity metric (Q1/3) for WGB and \toolname.}
\label{tab:robust1}
\setlength{\tabcolsep}{5pt}{
\resizebox{0.99\textwidth}{!}{
\begin{NiceTabular}{l|lcccccccccccccc}
\toprule
                              & \multirow{2}{*}{\textbf{Datasets}} & \multicolumn{2}{c}{\textbf{RBOVG}} & \multicolumn{4}{c}{\textbf{WGB}{\tt (BAR $\tau=0.5$)}}                  & \multicolumn{4}{c}{\textbf{\toolname} \textit{(S.T.A.)} {\tt (HMS $\tau=0.75$)}}            & \multicolumn{4}{c}{\textbf{\toolname} \textit{(S.T.M.)} {\tt (HMS $\tau=0.75$)}}            \\
\cmidrule(lr){3-4} \cmidrule(lr){5-8} \cmidrule(lr){9-12} \cmidrule(lr){13-16}
                              &    & OVA         & FPR         & OVA     & FPR    & \!\!$\mathcal{M}_i$(Q1/3)\!\!    & \!\!$\mathcal{M}_w$(Q1/3)\!\!    & OVA     & FPR   & \!\!$\mathcal{M}_i$(Q1/3)\!\!    & \!\!$\mathcal{M}_w$(Q1/3)\!\!    & OVA     & FPR   & \!\!$\mathcal{M}_i$(Q1/3)\!\!    & \!\!$\mathcal{M}_w$(Q1/3)\!\!    \\
\midrule
\multirow{6}{*}{\rotatebox{-90}{\textbf{Inductive}}}    & {\tt Cora}                         & 76.0      & 43.0      & 100.0 & 0.0  & 0.00/0.01 & 0.82/0.96 & 94.7  & 0.0 & 0.48/0.55 & 0.78/0.80 & 84.2  & 0.0 & 0.45/0.54 & 0.73/0.80 \\
                              & {\tt DBLP}                         & 93.5      & 13.0      & 100.0 & 0.0  & 0.00/0.00 & 0.98/1.00 & 100.0 & 0.0 & 0.47/0.55 & 0.95/0.98 & 58.2  & 0.0 & 0.45/0.52 & 0.66/0.73 \\
                              & {\tt Photo}                        & 58.0      & 84.0     & 92.0  & 0.0  & 0.00/0.01 & 0.63/0.83 & 84.2  & 0.0 & 0.47/0.54 & 0.75/0.84 & 100.0 & 0.0 & 0.45/0.55 & 1.00/1.00 \\
                              & {\tt CS}                           & 52.0      & 96.0      & 75.0  & 0.0  & 0.00/0.01 & 0.46/0.71 & 100.0 & 0.0 & 0.47/0.55 & 0.97/0.98 & 100.0 & 0.0 & 0.47/0.56 & 1.00/1.00 \\
                              & {\tt Physics}                      & 92.0      & 16.0      & 100.0 & 0.0  & 0.00/0.01 & 1.00/1.00 & 100.0 & 0.0 & 0.47/0.55 & 0.97/0.98 & 100.0 & 0.0 & 0.47/0.55 & 0.97/0.98 \\
                              & {\tt Blog}                         & 100.0     & 0.0       & 90.0  & 20.0 & 0.02/0.67 & 0.76/0.97 & 100.0 & 0.0 & 0.49/0.55 & 1.00/1.00 & 100.0 & 0.0 & 0.49/0.55 & 0.98/1.00 \\
\midrule
\midrule
\multirow{6}{*}{\rotatebox{-90}{\textbf{Transductive}}} & {\tt Cora}                         & 94.0      & 0.0       & 92.5  & 0.0  & 0.00/0.00 & 0.66/0.95 & 50.0  & 0.0 & 0.49/0.53 & 0.63/0.66 & 50.0  & 0.0 & 0.45/0.53 & 0.61/0.68 \\
                              & {\tt DBLP}                         & 100.0     & 0.0       & 100.0 & 0.0  & 0.01/0.04 & 0.86/0.94 & 100.0 & 0.0 & 0.45/0.55 & 0.92/0.97 & 78.6  & 0.0 & 0.45/0.55 & 0.72/0.80 \\
                              & {\tt Photo}                        & 90.0      & 0.0      & 77.5  & 0.0  & 0.00/0.02 & 0.87/0.99 & 89.5  & 0.0 & 0.47/0.55 & 0.75/0.85 & 100.0 & 0.0 & 0.45/0.54 & 0.97/1.00 \\
                              & {\tt CS}                           & 50.0      & 100.0     & 75.0  & 0.0  & 0.00/0.01 & 0.00/0.01 & 100.0 & 0.0 & 0.47/0.55 & 0.92/0.97 & 100.0 & 0.0 & 0.44/0.50 & 1.00/1.00 \\
                              & {\tt Physics}                      & 100.0     & 0.0       & 100.0 & 0.0  & 0.00/0.01 & 1.00/1.00 & 100.0 & 0.0 & 0.45/0.53 & 0.97/1.00 & 100.0 & 0.0 & 0.47/0.56 & 0.97/1.00 \\
                              & {\tt Blog}                         & 100.0     & 0.0       & 96.0  & 8.0  & 0.16/0.50 & 1.00/1.00 & 100.0 & 0.0 & 0.48/0.56 & 0.97/1.00 & 100.0 & 0.0 & 0.44/0.52 & 0.98/1.00 \\
\bottomrule
\end{NiceTabular}
}
}
\end{table*}

\begin{table*}[ht]\centering
\caption{Ownership verification accuracy (OVA) and false positive rate (FPR) of \toolname, WGB, and RBOVG after fine-tuning both the independently trained models and the watermarked (or fingerprinted) models with 200 epochs on the adversary graph. We additionally report the first and third quartiles of the similarity metric (Q1/3) for WGB and \toolname.}
\label{tab:robust2}
\setlength{\tabcolsep}{5pt}{
\resizebox{0.99\textwidth}{!}{
\begin{NiceTabular}{l|lcccccccccccccc}
\toprule
                              & \multirow{2}{*}{\textbf{Datasets}} & \multicolumn{2}{c}{\textbf{RBOVG}} & \multicolumn{4}{c}{\textbf{WGB}{\tt (BAR $\tau=0.5$)}}                 & \multicolumn{4}{c}{\textbf{\toolname} \textit{(S.T.A.)} {\tt (HMS $\tau=0.75$)}}            & \multicolumn{4}{c}{\textbf{\toolname} \textit{(S.T.M.)} {\tt (HMS $\tau=0.75$)}}                               \\
\cmidrule(lr){3-4} \cmidrule(lr){5-8} \cmidrule(lr){9-12} \cmidrule(lr){13-16}
                              &          & OVA          & FPR        & OVA     & FPR   &\!\!$\mathcal{M}_i$(Q1/3)\!\!    & \!\!$\mathcal{M}_w$(Q1/3)\!\!    & OVA     & FPR   & \!\!$\mathcal{M}_i$(Q1/3)\!\!    & \!\!$\mathcal{M}_w$(Q1/3)\!\!    & OVA     & FPR   & \!\!$\mathcal{M}_i$(Q1/3)\!\!    & \!\!$\mathcal{M}_w$(Q1/3)\!\!    \\
\midrule
\multirow{6}{*}{\rotatebox{-90}{\textbf{Inductive}}}    & {\tt Cora}     & 100.0      & 0.0      & 90.0  & 0.0  & 0.00/0.00 & 0.52/0.93 & 100.0 & 0.0 & 0.47/0.54 & 0.89/0.92 & 100.0 & 0.0 & 0.45/0.55 & 0.88/0.92 \\
                              & {\tt DBLP}     & 68.0       & 61.0     & 100.0 & 0.0  & 0.00/0.00 & 0.79/0.93 & 99.0  & 0.0 & 0.47/0.56 & 0.81/0.84 & 65.3  & 0.0 & 0.47/0.53 & 0.69/0.75 \\
                              & {\tt Photo}    & 100.0      & 0.0      & 100.0 & 0.0  & 0.00/0.01 & 0.68/0.78 & 100.0 & 0.0 & 0.47/0.52 & 0.89/0.93 & 100.0 & 0.0 & 0.45/0.56 & 1.00/1.00 \\
                              & {\tt CS}       & 85.0       & 2.0      & 100.0 & 0.0  & 0.00/0.00 & 0.93/0.94 & 100.0 & 0.0 & 0.45/0.53 & 0.98/0.98 & 100.0 & 0.0 & 0.47/0.53 & 0.98/1.00 \\
                              & {\tt Physics}  & 100.0      & 0.0      & 100.0 & 0.0  & 0.00/0.04 & 0.94/0.97 & 100.0 & 0.0 & 0.45/0.52 & 0.97/0.98 & 100.0 & 0.0 & 0.44/0.53 & 0.91/0.95 \\
                              & {\tt Blog}     & 100.0      & 0.0      & 73.0  & 54.0 & 0.38/0.61 & 1.00/1.00 & 100.0 & 0.0 & 0.42/0.55 & 1.00/1.00 & 100.0 & 0.0 & 0.48/0.55 & 1.00/1.00 \\
\midrule
\midrule
\multirow{6}{*}{\rotatebox{-90}{\textbf{Transductive}}} & {\tt Cora}     & 68.0       & 0.0      & 100.0 & 0.0  & 0.00/0.01 & 0.97/1.00 & 81.6  & 0.0 & 0.46/0.53 & 0.71/0.77 & 86.8  & 0.0 & 0.45/0.55 & 0.73/0.78 \\
                              & {\tt DBLP}     & 50.0       & 0.0      & 100.0 & 0.0  & 0.00/0.01 & 0.83/1.00 & 75.5  & 0.0 & 0.47/0.55 & 0.70/0.80 & 62.2  & 0.0 & 0.47/0.56 & 0.66/0.73 \\
                              & {\tt Photo}    & 58.0       & 0.0      & 100.0 & 0.0  & 0.00/0.01 & 0.77/0.86 & 100.0 & 0.0 & 0.45/0.55 & 0.94/0.97 & 100.0 & 0.0 & 0.45/0.55 & 1.00/1.00 \\
                              & {\tt CS}       & 49.0       & 2.0      & 100.0 & 0.0  & 0.00/0.01 & 0.98/1.00 & 100.0 & 0.0 & 0.47/0.56 & 0.97/0.98 & 100.0 & 0.0 & 0.45/0.52 & 0.98/1.00 \\
                              & {\tt Physics}  & 100.0      & 0.0      & 100.0 & 0.0  & 0.00/0.00 & 0.83/0.95 & 100.0 & 0.0 & 0.48/0.56 & 0.98/1.00 & 100.0 & 0.0 & 0.47/0.53 & 0.89/0.94 \\
                              & {\tt Blog}     & 50.0       & 0.0      & 81.0  & 38.0 & 0.33/0.59 & 1.00/1.00 & 100.0 & 0.0 & 0.48/0.55 & 1.00/1.00 & 100.0 & 0.0 & 0.46/0.53 & 1.00/1.00 \\
\bottomrule
\end{NiceTabular}
}
}
\end{table*}

\begin{itemize}[leftmargin=*]
    \item \textit{Pruning}: The adversary performs L1-norm pruning by zeroing out the neurons with the smallest L1 norms. We report changes in HMS and TAC as the pruning ratio increases from 0\% to 50\%, 60\%, 70\%, 80\%, and 90\%.
    \item \textit{Fine-tuning}: The adversary fine-tunes the watermarked model on the adversary graph using the Adam optimizer with a learning rate of 5e-5. We report the variations in HMS and TAC as fine-tuning epochs increases.
    \item \textit{Ownership collision}: Independently trained models may be pruned for compression. The adversary graph may be acquired inadvertently by benign users for fine-tuning. We apply pruning (70\% ratio) and fine-tuning (200 epochs) for both independently trained and watermarked models to assess whether false positives arise.
    \item \textit{Overwriting:} A watermarked model may undergo subsequent watermark embedding by the adversary. The new watermark potentially overwrites the original watermark. We apply \toolname in \textit{S.T.M.} to embed a new watermark in the watermarked model and assess the detectability of the original watermark. 
    \item \textit{Model extraction attacks}: The adversary uses the adversary graph to query the watermarked model's API and obtain the prediction probabilities. The adversary then trains a surrogate model on these samples. The surrogate model is the SAGE architecture and is trained by the Adam optimizer with a learning rate of 1e-4 for 1500 epochs, with the KL divergence serving as the objective loss function. 
\end{itemize}

\begin{figure}[!t]
    \centering
    \includegraphics[page=1, width=0.49\linewidth]{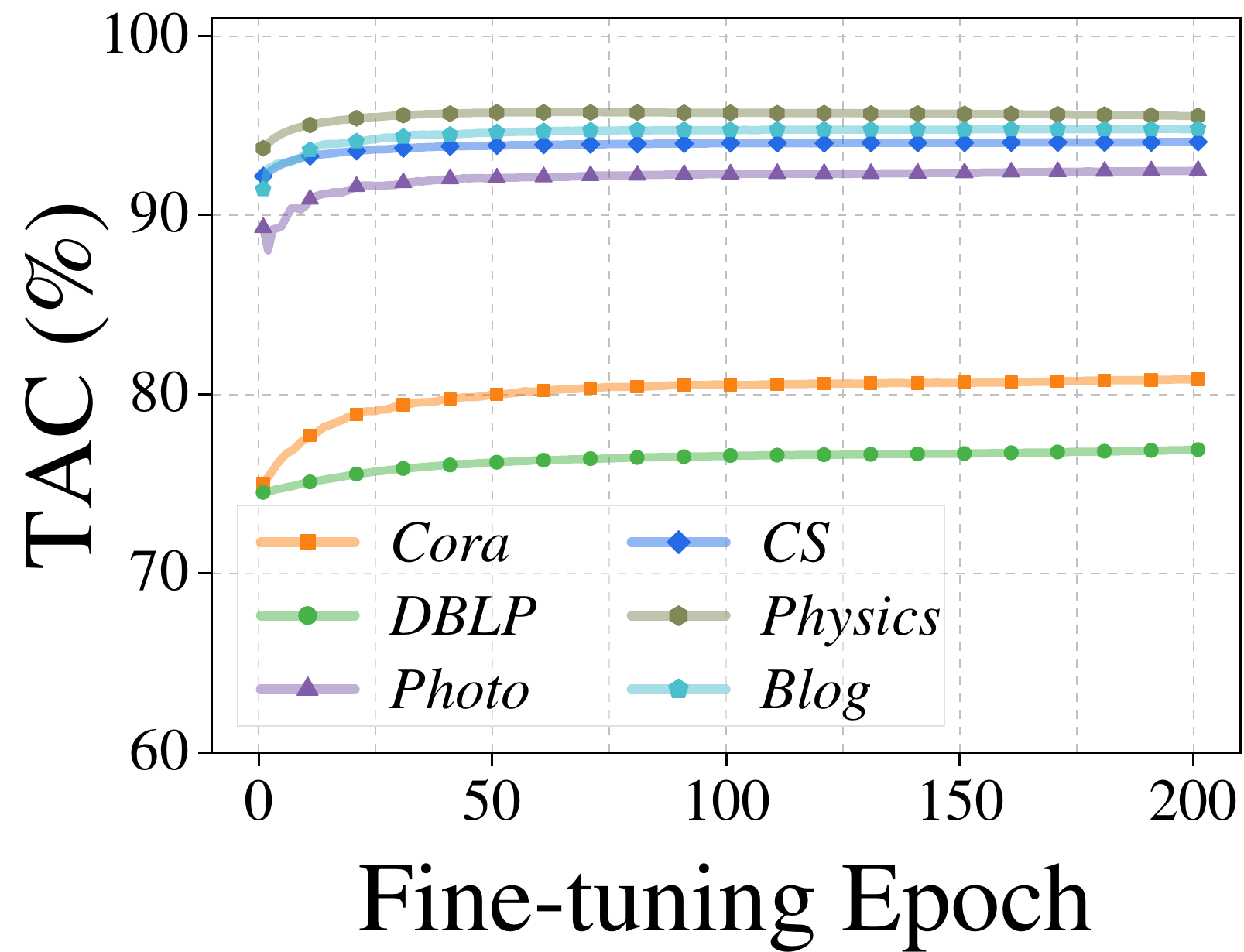}
    \includegraphics[page=2, width=0.49\linewidth]{fig/Fine-tuning.pdf}
    \caption{Test accuracy (TAC) and Hamming similarity (HMS) with increasing fine-tuning epoch. This figure depicts \toolname in \textit{S.T.A.} under the inductive paradigm.}
    \label{fig:fine-tuning}
    \vspace{-2mm}
\end{figure}

\noindent \textbf{Result Analysis.} 
\noindent \textit{Pruning.} Figure~\ref{fig:pruning} illustrates the impact of varying pruning ratios on the Hamming similarity (HMS) and test accuracy (TAC) of \toolname. In general, both metrics degrade as the pruning ratio increases. {\tt Photo} exhibits limited robustness, with HMS dropping below the similarity threshold (0.75) at an 80\% pruning ratio. However, its TAC accompany a substantial concurrent decline to less than 50\%. {\tt CS} demonstrates strong robustness, maintaining a high HMS even at an 80\% pruning ratio when TAC falls to 50\%. The results demonstrate that the adversary cannot disrupt the watermark without compromising model performance.

\vspace{1mm}
\noindent \textit{Fine-tuning.} Figure~\ref{fig:fine-tuning} depicts Hamming similarity (HMS) degradation with increasing fine-tuning epochs. In most cases, the HMS fluctuates at begin and then stabilizes. {\tt DBLP} shows the weakest robustness, yet its HMS exceeds the 0.75 similarity threshold after 200 epochs. The weak robustness of {\tt DBLP} is due to the limited class. These results indicate that fine-tuning improves model performance while inflicting minor watermark damage, without impairing ownership verification.

\vspace{1mm}
\noindent \textit{Ownership collision.} 
Ownership collision refers to the misclassification of independently trained models as copies, which undermines the credibility of ownership verification.
Tables~\ref{tab:robust1} and~\ref{tab:robust2} report the ownership verification results after applying pruning or fine-tuning to both independently trained models and watermarked (or fingerprinted) models.
\toolname demonstrates the strongest robustness. After pruning, \toolname \textit{(S.T.A.)} achieves 100\% OVA on eight datasets, WGB on five, and RBOVG on four.
Critically, \toolname maintains zero FPR in all cases, ensuring independently trained models are never misidentified as unauthorized copies.

\begin{table*}[ht]
\centering
\caption{Test accuracy (TAC) and hamming similarity (HMS) of watermarked models ($\mathcal{M}_w$) after overwriting. Values after $\downarrow\uparrow$ indicate the changes compared to before overwriting.}
\label{tab:overwriting}
\setlength{\tabcolsep}{9pt}{
\resizebox{0.99\textwidth}{!}{
\begin{NiceTabular}{l|lcccccccccc}
\toprule
& \multirow{2}{*}{\textbf{Datasets}} & \multicolumn{5}{c}{\textbf{\toolname} \textit{(S.T.A.)} {\tt (HMS $\tau=0.75$)}}                                                                                       & \multicolumn{5}{c}{\textbf{\toolname} \textit{(S.T.M.)} {\tt (HMS $\tau=0.75$)}}                                                    \\
\cmidrule(lr){3-7} \cmidrule(lr){8-12}
 &        & TAC          & OVA & FPR & Q1 & Q3 & TAC         & OVA     & FPR   & Q1 &   Q3 \\
\midrule
\multirow{6}{*}{\rotatebox{-90}{\textbf{Inductive}}}    & {\tt Cora}                         & 74.67 {\tiny (\textcolor{red}{$\downarrow$0.32})}  & 100.0                 & 0.0                   & 0.92 {\tiny (\textcolor{red}{$\downarrow$0.01})}             & 0.98 {\tiny (\textcolor{red}{$\downarrow$0.02})}             & 76.47 {\tiny (\textcolor{red}{$\downarrow$0.57})} & 100.0 & 0.0 & 1.00 {\tiny (\textcolor{red}{$\downarrow$0.04})}             & 1.00 {\tiny (\textcolor{red}{$\downarrow$0.02})}             \\
                              & {\tt DBLP}                         & 74.33  {\tiny (\textcolor{red}{$\downarrow$0.18})} & 100.0                 & 0.0                   & 1.00 {\tiny (\textcolor{red}{$\downarrow$0.03})}             & 1.00 {\tiny (\textcolor{red}{$\downarrow$0.00})}             & 75.40 {\tiny (\textcolor{red}{$\downarrow$0.11})} & 100.0 & 0.0 & 1.00 {\tiny (\textcolor{red}{$\downarrow$0.03})}             & 1.00 {\tiny (\textcolor{red}{$\downarrow$0.00})}             \\
                              & {\tt Photo}                        & 83.49 {\tiny (\textcolor{red}{$\downarrow$5.81})} & 100.0                 & 0.0                   & 0.88 {\tiny (\textcolor{red}{$\downarrow$0.02})}             & 0.93 {\tiny (\textcolor{red}{$\downarrow$0.03})}             & 86.15 {\tiny (\textcolor{red}{$\downarrow$3.61})} & 100.0 & 0.0 & 1.00 {\tiny (\textcolor{red}{$\downarrow$0.00})}             & 1.00 {\tiny (\textcolor{red}{$\downarrow$0.00})}             \\
                              & {\tt CS}                           & 91.55 {\tiny (\textcolor{red}{$\downarrow$0.62})} & 100.0                 & 0.0                   & 0.98 {\tiny (\textcolor{red}{$\downarrow$0.02})}             & 0.98 {\tiny (\textcolor{red}{$\downarrow$0.00})}             & 93.70 {\tiny (\textcolor{red}{$\downarrow$0.04})} & 100.0 & 0.0 & 1.00 {\tiny (\textcolor{red}{$\downarrow$0.00})}             & 1.00 {\tiny (\textcolor{red}{$\downarrow$0.00})}             \\
                              & {\tt Physics}                      & 93.59 {\tiny (\textcolor{red}{$\downarrow$0.15})} & 100.0                 & 0.0                   & 0.98 {\tiny (\textcolor{red}{$\downarrow$0.00})}             & 1.00 {\tiny (\textcolor{red}{$\downarrow$0.00})}             & 95.87 {\tiny (\textcolor{red}{$\downarrow$0.02})} & 100.0 & 0.0 & 1.00 {\tiny (\textcolor{red}{$\downarrow$0.00})}             & 1.00 {\tiny (\textcolor{red}{$\downarrow$0.00})}             \\
                              & {\tt Blog}                         & 91.32 {\tiny (\textcolor{red}{$\downarrow$0.12})} & 100.0                 & 0.0                   & 1.00 {\tiny (\textcolor{red}{$\downarrow$0.00})}             & 1.00 {\tiny (\textcolor{red}{$\downarrow$1.00})}             & 93.83 {\tiny (\textcolor{red}{$\downarrow$0.06})} & 100.0 & 0.0 & 1.00 {\tiny (\textcolor{red}{$\downarrow$0.00})}             & 1.00 {\tiny (\textcolor{red}{$\downarrow$1.00})}             \\
\midrule
\midrule
\multirow{6}{*}{\rotatebox{-90}{\textbf{Transductive}}} & {\tt Cora}                         & 78.50 {\tiny (\textcolor{red}{$\downarrow$0.17})} & 100.0                 & 0.0                   & 0.95 {\tiny (\textcolor{red}{$\downarrow$0.02})}             & 0.97 {\tiny (\textcolor{red}{$\downarrow$0.00})}             & 80.87 {\tiny (\textcolor{red}{$\downarrow$0.34})} & 100.0 & 0.0 & 0.97 {\tiny (\textcolor{red}{$\downarrow$0.05})}             & 1.00 {\tiny (\textcolor{red}{$\downarrow$0.04})}             \\
                              & {\tt DBLP}                         & 81.48 {\tiny (\textcolor{red}{$\downarrow$0.18})} & 100.0                 & 0.0                   & 1.00 {\tiny (\textcolor{red}{$\downarrow$0.05})}             & 1.00 {\tiny (\textcolor{red}{$\downarrow$0.02})}             & 83.32 {\tiny (\textcolor{red}{$\downarrow$0.33})} & 100.0 & 0.0 & 0.97 {\tiny (\textcolor{red}{$\downarrow$0.03})}             & 0.98 {\tiny (\textcolor{red}{$\downarrow$0.03})}             \\
                              & {\tt Photo}                        & 85.56 {\tiny (\textcolor{red}{$\downarrow$7.59})} & 100.0                 & 0.0                   & 0.95 {\tiny (\textcolor{red}{$\downarrow$0.05})}             & 0.98 {\tiny (\textcolor{red}{$\downarrow$0.04})}             & 89.71 {\tiny (\textcolor{red}{$\downarrow$3.85})} & 100.0 & 0.0 & 1.00 {\tiny (\textcolor{red}{$\downarrow$0.02})}             & 1.00 {\tiny (\textcolor{red}{$\downarrow$0.00})}             \\
                              & {\tt CS}                           & 94.11 {\tiny (\textcolor{red}{$\downarrow$0.11})} & 100.0                 & 0.0                   & 1.00 {\tiny (\textcolor{red}{$\downarrow$0.02})}             & 1.00 {\tiny (\textcolor{red}{$\downarrow$0.00})}             & 95.50 {\tiny (\textcolor{red}{$\downarrow$0.12})} & 100.0 & 0.0 & 1.00 {\tiny (\textcolor{red}{$\downarrow$0.00})}             & 1.00 {\tiny (\textcolor{red}{$\downarrow$0.00})}             \\
                              & {\tt Physics}                      & 94.61 {\tiny (\textcolor{red}{$\downarrow$0.21})} & 100.0                 & 0.0                   & 1.00 {\tiny (\textcolor{red}{$\downarrow$0.00})}             & 1.00 {\tiny (\textcolor{red}{$\downarrow$0.00})}             & 96.98 {\tiny (\textcolor{red}{$\downarrow$0.05})} & 100.0 & 0.0 & 1.00 {\tiny (\textcolor{red}{$\downarrow$0.00})}             & 1.00 {\tiny (\textcolor{red}{$\downarrow$0.00})}             \\
                              & {\tt Blog}                         & 92.67  {\tiny (\textcolor{blue}{$\uparrow$0.02})} & 100.0                 & 0.0                   & 1.00 {\tiny (\textcolor{red}{$\downarrow$0.00})}             & 1.00 {\tiny (\textcolor{red}{$\downarrow$0.00})}             & 95.96 {\tiny (\textcolor{red}{$\downarrow$0.16})} & 100.0 & 0.0 & 1.00 {\tiny (\textcolor{red}{$\downarrow$0.00})}             & 1.00 {\tiny (\textcolor{red}{$\downarrow$0.00})} \\
\bottomrule
\end{NiceTabular}
}
}
\end{table*}

\begin{table*}[ht]
\centering
\caption{Test accuracy (TAC), ownership verification accuracy (OVA), and false positive rate (FPR) of watermarked (or fingerprinted) models and their corresponding surrogate models using the SAGE architecture. }
\label{tab:extraction}
\setlength{\tabcolsep}{4pt}{
\resizebox{1.0\textwidth}{!}{
\centering
\begin{NiceTabular}{l|lccccccccccccccc}
\toprule
  & \textbf{\multirow{2}{*}{Datasets}} & \multicolumn{3}{c}{\textbf{RBOVG}}      & \multicolumn{4}{c}{\textbf{WGB} {\tt (BAR $\tau=0.5$)}}                    & \multicolumn{4}{c}{\textbf{\toolname} \textit{(S.T.A.)} {\tt (HMS $\tau=0.75$)}}                & \multicolumn{4}{c}{\textbf{\toolname} \textit{(S.T.M.)} {\tt (HMS $\tau=0.75$)}}               \\
\cmidrule(lr){3-5} \cmidrule(lr){6-9} \cmidrule(lr){10-13} \cmidrule(lr){14-17}
  &    & TAC          & OVA     & FPR   & TAC          & OVA    & FPR   & Q1/3       & TAC          & OVA     & FPR   & Q1/3       & TAC          & OVA    & FPR   & Q1/3       \\
\midrule
\multirow{6}{*}{\rotatebox{-90}{\textbf{Inductive}}}    & {\tt Cora}                         & 65.23 {\tiny (\textcolor{red}{$\downarrow$5.87})}  & 100.0 & 0.0 & 68.97 {\tiny (\textcolor{red}{$\downarrow$2.68})}  & 50.0  & 0.0 & 0.21/0.24 & 69.15 {\tiny (\textcolor{red}{$\downarrow$5.84})}  & 50.0  & 0.0 & 0.48/0.61 & 70.97 {\tiny (\textcolor{red}{$\downarrow$6.08})}  & 50.0 & 0.0 & 0.48/0.54 \\
                              & {\tt DBLP}                         & 70.60 {\tiny (\textcolor{red}{$\downarrow$1.64})}  & 100.0 & 0.0 & 74.56 {\tiny (\textcolor{blue}{$\uparrow$1.48})} & 50.0  & 0.0 & 0.10/0.29 & 74.81 {\tiny (\textcolor{blue}{$\uparrow$0.30})} & 91.0  & 0.0 & 0.75/0.83 & 75.69 {\tiny (\textcolor{blue}{$\uparrow$0.18})} & 50.0 & 0.0 & 0.53/0.59 \\
                              & {\tt Photo}                        & 90.94 {\tiny (\textcolor{blue}{$\uparrow$0.71})} & 100.0 & 0.0 & 87.67 {\tiny (\textcolor{red}{$\downarrow$4.40})}  & 50.0  & 0.0 & 0.02/0.03 & 88.78 {\tiny (\textcolor{red}{$\downarrow$0.53})}  & 81.0  & 0.0 & 0.74/0.84 & 93.15 {\tiny (\textcolor{red}{$\downarrow$0.41})}  & 50.0 & 0.0 & 0.57/0.65 \\
                              & {\tt CS}                           & 89.71 {\tiny (\textcolor{red}{$\downarrow$1.34})}  & 100.0 & 0.0 & 92.36 {\tiny (\textcolor{red}{$\downarrow$1.45})}  & 50.0  & 0.0 & 0.00/0.00 & 90.93 {\tiny (\textcolor{red}{$\downarrow$1.24})}  & 100.0 & 0.0 & 0.84/0.89 & 92.22 {\tiny (\textcolor{red}{$\downarrow$1.52})}  & 50.0 & 0.0 & 0.57/0.65 \\
                              & {\tt Physics}                      & 94.14 {\tiny (\textcolor{blue}{$\uparrow$0.50})} & 100.0 & 0.0 & 95.43 {\tiny (\textcolor{red}{$\downarrow$0.07})}  & 50.0  & 0.0 & 0.00/0.00 & 94.32 {\tiny (\textcolor{blue}{$\uparrow$0.58})} & 100.0 & 0.0 & 0.88/0.92 & 95.58 {\tiny (\textcolor{red}{$\downarrow$0.31})}  & 50.0 & 0.0 & 0.47/0.56 \\
                              & {\tt Blog}                         & 79.50 {\tiny (\textcolor{red}{$\downarrow$10.27})} & 100.0 & 0.0 & 81.27 {\tiny (\textcolor{red}{$\downarrow$12.44})} & 100.0 & 0.0 & 0.86/0.95 & 88.38 {\tiny (\textcolor{red}{$\downarrow$3.05})}  & 100.0 & 0.0 & 0.94/0.96 & 90.02 {\tiny (\textcolor{red}{$\downarrow$3.88})}  & 50.0 & 0.0 & 0.48/0.58 \\
\midrule
\midrule
\multirow{6}{*}{\rotatebox{-90}{\textbf{Transductive}}} & {\tt Cora}                         & 78.87 {\tiny (\textcolor{red}{$\downarrow$2.70})}  & 100.0 & 0.0 & 80.44 {\tiny (\textcolor{red}{$\downarrow$0.17})}  & 50.0  & 0.0 & 0.01/0.03 & 77.89 {\tiny (\textcolor{red}{$\downarrow$0.78})}  & 50.0  & 0.0 & 0.59/0.64 & 79.67 {\tiny (\textcolor{red}{$\downarrow$1.55})}  & 50.0 & 0.0 & 0.46/0.56 \\
                              & {\tt DBLP}                         & 79.33 {\tiny (\textcolor{red}{$\downarrow$0.26})}  & 100.0 & 0.0 & 81.20 {\tiny (\textcolor{red}{$\downarrow$0.36})}  & 50.0  & 0.0 & 0.00/0.05 & 81.05 {\tiny (\textcolor{red}{$\downarrow$0.61})}  & 100.0 & 0.0 & 0.88/0.94 & 82.84 {\tiny (\textcolor{red}{$\downarrow$0.81})}  & 50.0 & 0.0 & 0.52/0.58 \\
                              & {\tt Photo}                        & 93.16 {\tiny (\textcolor{red}{$\downarrow$0.52})}  & 100.0 & 0.0 & 93.64 {\tiny (\textcolor{red}{$\downarrow$1.48})}  & 50.0  & 0.0 & 0.01/0.01 & 93.43 {\tiny (\textcolor{blue}{$\uparrow$0.28})} & 100.0 & 0.0 & 0.81/0.86 & 88.87 {\tiny (\textcolor{red}{$\downarrow$0.89})}  & 50.0 & 0.0 & 0.52/0.60 \\
                              & {\tt CS}                           & 91.64 {\tiny (\textcolor{red}{$\downarrow$1.07})}  & 100.0 & 0.0 & 92.73 {\tiny (\textcolor{red}{$\downarrow$2.59})}  & 75.0  & 0.0 & 0.00/0.01 & 92.73 {\tiny (\textcolor{red}{$\downarrow$1.49})}  & 100.0 & 0.0 & 0.81/0.89 & 93.39 {\tiny (\textcolor{red}{$\downarrow$2.22})}  & 50.0 & 0.0 & 0.44/0.53 \\
                              & {\tt Physics}                      & 95.54 {\tiny (\textcolor{blue}{$\uparrow$1.56})} & 100.0 & 0.0 & 95.97 {\tiny (\textcolor{red}{$\downarrow$1.01})}  & 50.0  & 0.0 & 0.00/0.00 & 94.63 {\tiny (\textcolor{red}{$\downarrow$0.19})}  & 100.0 & 0.0 & 0.92/0.97 & 96.06 {\tiny (\textcolor{red}{$\downarrow$0.96})}  & 50.0 & 0.0 & 0.45/0.55 \\
                              & {\tt Blog}                         & 75.77 {\tiny (\textcolor{red}{$\downarrow$10.90})} & 100.0 & 0.0 & 93.00 {\tiny (\textcolor{red}{$\downarrow$2.86})}  & 84.0  & 0.0 & 0.42/0.87 & 91.53 {\tiny (\textcolor{red}{$\downarrow$1.13})}  & 100.0 & 0.0 & 0.95/0.98 & 92.73 {\tiny (\textcolor{red}{$\downarrow$3.39})}  & 50.0 & 0.0 & 0.52/0.56 \\
\bottomrule
\end{NiceTabular}
}
}
\end{table*}

\begin{figure}[!t]
    \centering
    \includegraphics[page=1, width=0.49\linewidth]{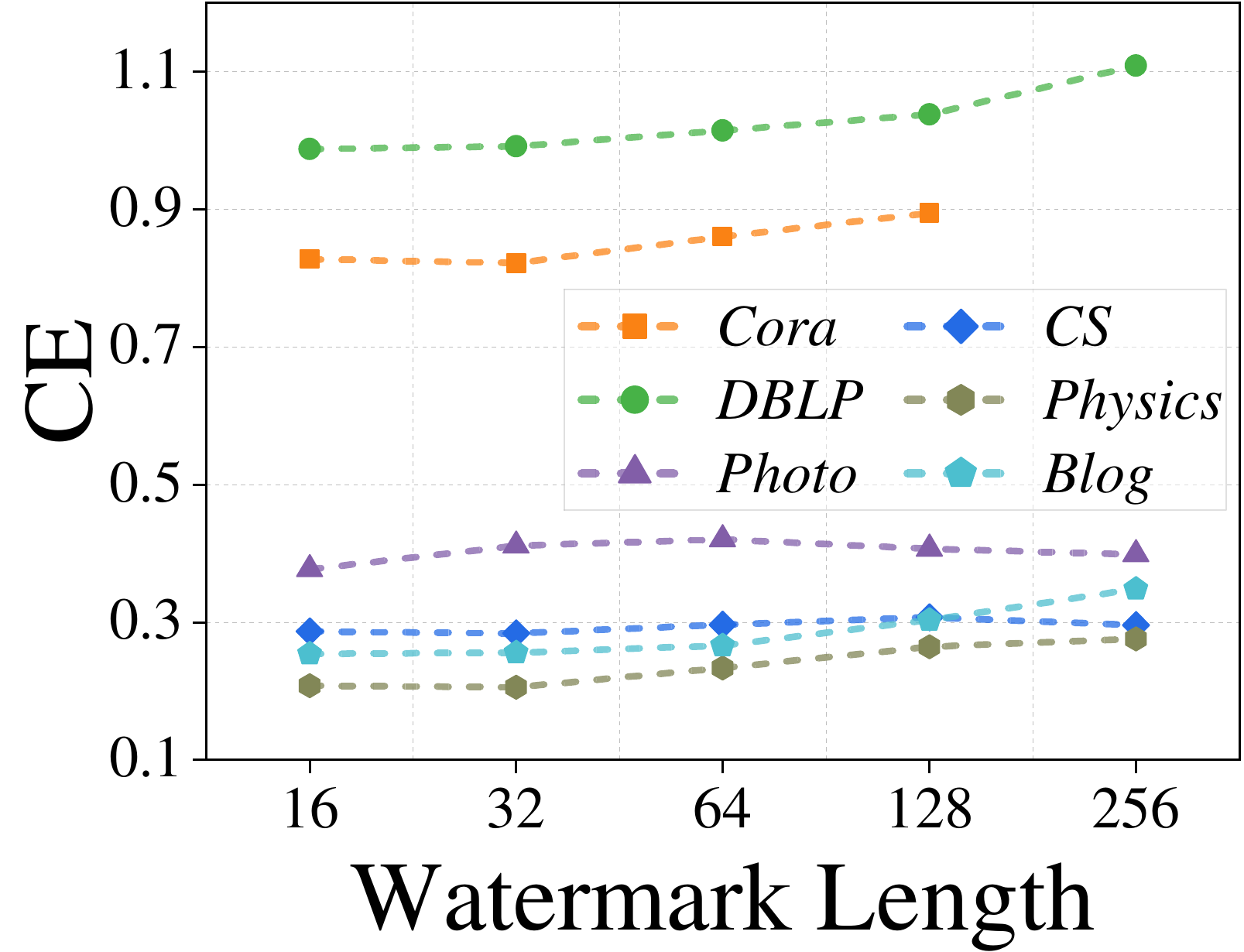}
    \includegraphics[page=2, width=0.49\linewidth]{fig/Multibit.pdf}
    \caption{The primacy task CE on the test graph and the watermark task BCE after fine-tuning. This figure depicts \toolname in \textit{S.T.A.} under the inductive paradigm.}
    \label{fig:multibit1}
    \vspace{-2mm}
\end{figure}

\textit{Overwriting.} 
Table~\ref{tab:overwriting} reports the changes in TAC and HMS before and after watermark overwriting. We observe that watermark overwriting leads to a decrease in both TAC and HMS. However, the original watermark maintains sufficiently high HMS, achieving 100\% OVA in all cases. These results indicate that using \toolname to embed a new watermark does not compromise the original watermark.
It should be noted that the watermarked model now contains two distinct watermarks, a situation commonly encountered in ownership verification. This issue can be addressed by registering the encrypted watermark with a third party (e.g., an intellectual property authority or a blockchain system) along with a timestamp~\cite{waheed2024grove}. The watermark with the later timestamp will not be recognized as proof of ownership.

\textit{Model Extraction}
Table~\ref{tab:extraction} presents the results of the ownership verification for the surrogate model obtained via model extraction attacks.
In this table, we observe that \toolname in \textit{S.T.A.} achieves high OVA and 0 FPR on most models except {\tt Cora}, indicating that the watermark remains detectable in most surrogate models. These results confirm that the watermark embedded in the watermarked model can be inherited by the surrogate models. Although \toolname in \textit{S.T.A.} is slightly less than RBOVG against model extraction attacks, it remains effective in most scenarios. WGB and \toolname in \textit{S.T.M.} fail against model extraction attacks. The surrogate model retains test accuracy comparable to that of the watermarked model on the primary task without the watermark. This is because WGB and \toolname in \textit{S.T.M.} only perturb the decision boundaries, whereas \toolname in \textit{S.T.A.} induces a global alteration of the decision space distribution. This whole alteration can be inherited by surrogate models.

\noindent \textbf{Answers to RQ2:} For \toolname, the adversary cannot remove the watermark by pruning or fine-tuning without severely degrading model performance. \toolname in \textit{S.T.A.} demonstrates robustness against model extraction attacks, with the watermark detectable in surrogate models in most cases. Importantly, \toolname consistently exhibits zero false positives.

\subsection{Impact of Hyper-parameters (RQ3)}
\noindent \textbf{Experiment Design.} The watermark bit string length $N_w$ and the judgment threshold $\tau$ are two key hyperparameters in \toolname. We discuss them as follows. 

\begin{itemize}[leftmargin=*]
    \item \textit{Watermark length $N_w$:} In prior experiments, we fixed $N_w=64$. In this part, we assess the fidelity and robustness of \toolname for $N \in \{16, 32, 64, 128, 256\}$ bits. Fidelity is measured by the primacy task cross-entropy (CE) on the test graph, while robustness is evaluated using the watermark task binary cross-entropy (BCE) of the watermarked model after fine-tuning. We do not use TAC and HMS as they fail to capture subtle variations in performance.
    \item \textit{Judgment threshold $\tau$:} In prior experiments, we fixed $\tau=0.75$, which is the midpoint of Hamming similarity. We discuss the selection of $\tau$ and show that the watermark collision probability follows a normal distribution.
\end{itemize}

\begin{figure}[t]
    \centering
    \includegraphics[page=1, width=0.49\linewidth]{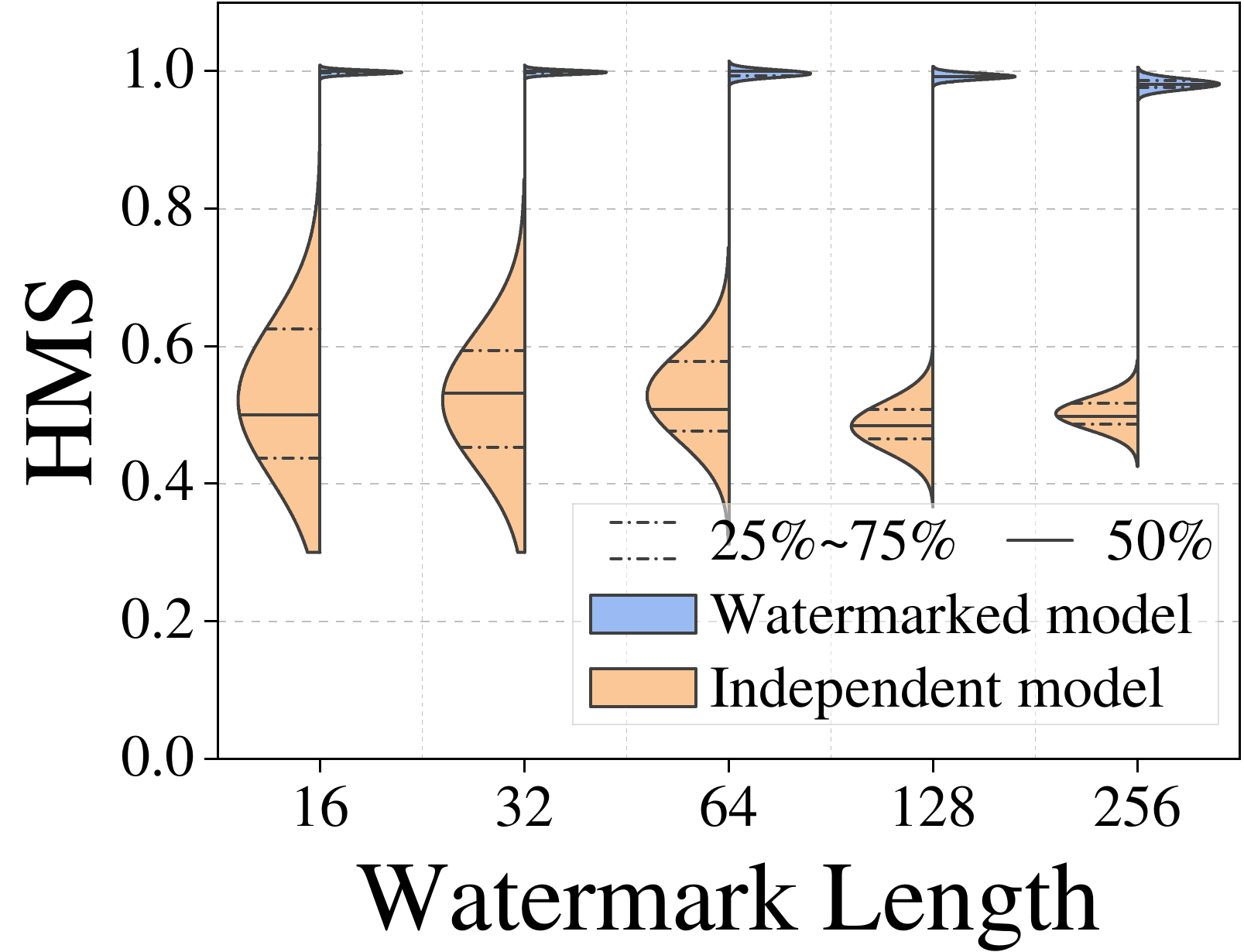}
    \includegraphics[page=2, width=0.49\linewidth]{fig/Multibit2.pdf}
    \caption{Violin plot of {\tt DBLP} with different watermark lengths. This figure depicts \toolname in \textit{S.T.A.}.}
    \label{fig:multibit2}
    \vspace{-2mm}
\end{figure}

\noindent \textbf{Result Analysis.}
As shown in Figure \ref{fig:multibit1}, both the primacy task cross-entropy (CE) and the watermark task binary cross-entropy (BCE) increase with the length of the watermark $N_w$. This indicates that longer watermarks degrade the fidelity and robustness of \toolname. However, shorter watermarks are not necessarily preferable, as they lead to a narrower watermark collision space, requiring higher verification thresholds to mitigate false positives. As shown in Figure~\ref{fig:multibit2}, shorter watermark lengths yield a narrower collision space between independently trained and watermarked models. When $N_w = 16$, the maximum HMS of independently trained models closely approaches the minimum HMS of the watermarked models. Thus, the watermark length must be long enough to mitigate the risks of false positives.

\begin{figure}[t]
    \centering
    \includegraphics[page=1, width=0.49\linewidth]{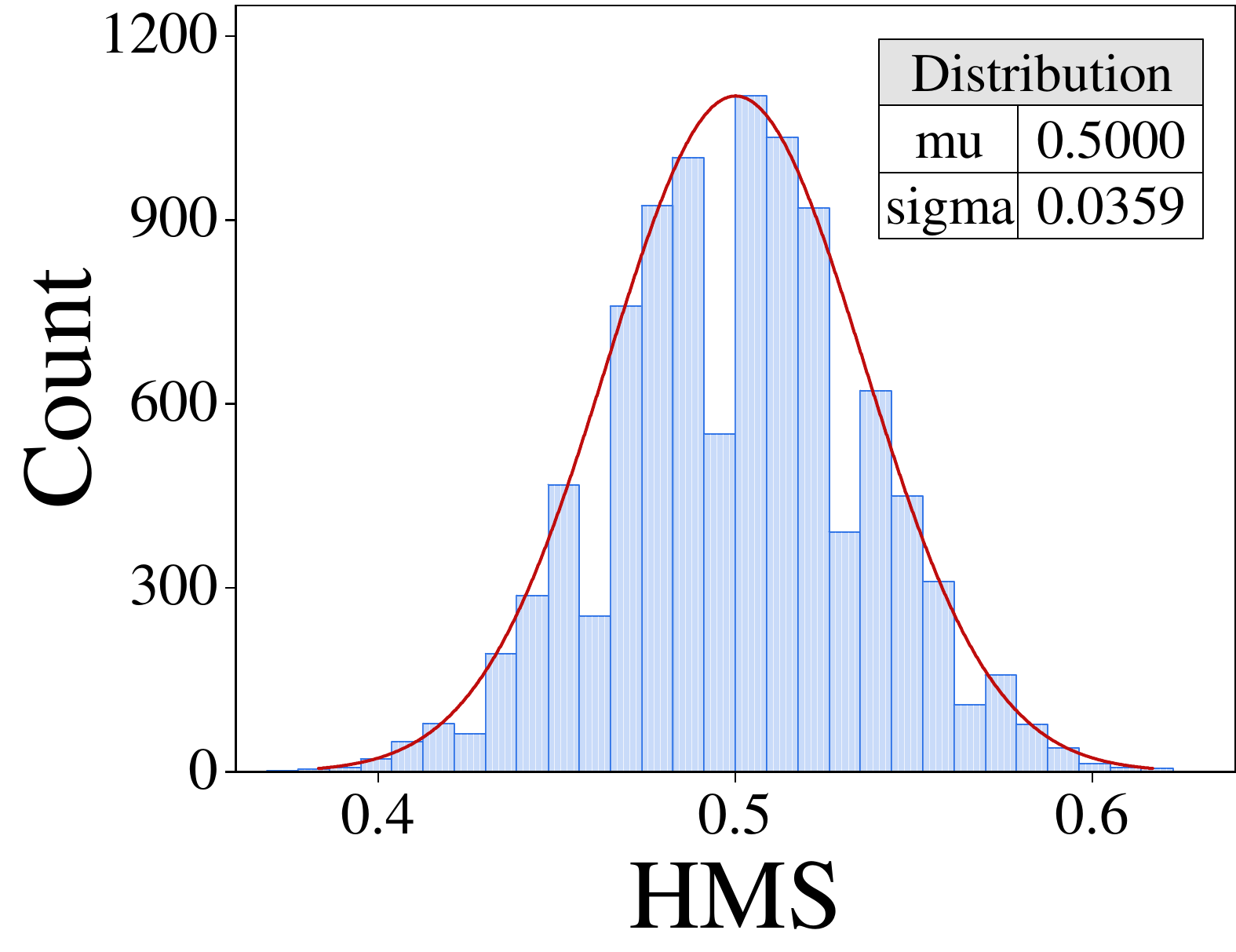}
    \includegraphics[page=2, width=0.49\linewidth]{fig/Collision.pdf}
    \caption{HMS between the target watermark and the extracted watermark from another watermarked model without the target watermark. This figure depicts {\tt Physics} in \toolname in \textit{S.T.A.} under the inductive paradigm.}
    \label{fig:collision}
    \vspace{-1mm}
\end{figure}

Figure \ref{fig:collision} illustrates the HMS from another watermarked model without the target watermark. The results show that the watermark collisions of \toolname follow a normal distribution centered at 0.5 with a standard deviation of approximately 0.35, consistent with our theoretical expectations. Therefore, the judgment threshold $\tau$ can be determined from a normal distribution and appropriately lowered when watermarked models are few or watermark strings are long. For example, setting $\tau = 0.7$ for Table\ref{tab:extraction} can achieve 100\% OVA for {\tt DBLP} and {\tt Photo} in \toolname \textit{(S.T.A.)}.

\noindent \textbf{Answers to RQ3:} A longer watermark length $N_w$ makes poorer fidelity and robustness but wider collision space. The ownership collision of \toolname follows a normal distribution. Higher $\tau$ can ensure zero FPR but diminish OVA.





\section{Limitation of \toolname}
\toolname achieves a backdoor-free, multi-bit watermarking scheme in the black-box setting for node-level GNNs. However, its applicability is inherently limited to GNNs, as LDDE is a property unique to graph-structured data. Consequently, \toolname is not applicable to conventional neural networks that operate on Euclidean data.
Furthermore, \toolname does not support graph-level GNNs in black-box settings. These models employ graph pooling operations to aggregate node embeddings into a single graph-level representation. Without observable edge structures in the outputs, \toolname cannot be achieved in black-box settings.

\toolname in \textit{S.T.A.} requires enough edges to carry watermarks, which is readily met in practice. When the edge count is insufficient for long watermarks, \toolname in \textit{S.T.M.} can serve as an effective alternative. Although \toolname in \textit{S.T.A.} exhibits robustness against model extraction attacks, its effectiveness diminishes on small-scale datasets (e.g., {\tt Cora}). Moreover, sufficiently long watermark bit strings are necessary to mitigate ownership collisions. We consider improving robustness against model extraction on small datasets and further increasing watermark capacity as future work.

\section{Conclusions}
This paper presents \toolname, a novel backdoor-free and multi-bit black-box watermarking paradigm for node-level GNNs. We introduce the insights of \textit{Layer-wise Distance Difference on an Edge (LDDE)} and prove LDDE suits for watermarks through theoretical analysis and experiments. 
\toolname takes the LDDE signs of the selected edges as a watermark rather than the backdoor mechanism, eliminating the risks of backdoor attacks. 
By modifying the LDDE values of each selected edge for respective target signs (positive or negative), \toolname supports multi-bit capacity, where each watermarked model is embedded with a unique watermark bit string while sharing the same trigger graph. This design enables ownership verification by only querying the suspect model once even if multiple watermarked models exist, making the process economical and hardly detectable.
To address practical ownership transfer scenarios where the new owner possesses only the original model without the training graph, we propose two embedding strategies: the setting that the training graph is available \textit{(S.T.A.)} and the setting that the training graph is missing \textit{(S.T.M.)}. Extensive experiments demonstrate the effectiveness, fidelity, robustness, and low overhead of \toolname.



\bibliographystyle{IEEEtran}
\bibliography{references}

\end{document}